\documentclass[10pt]{iopart}

\usepackage{iopams}  
\usepackage{graphicx}% Include figure files

\newcommand{\bol}[1]{\boldsymbol #1}
\def\rnum#1{\expandafter{%
\romannumeral #1}}
\def\Rnum#1{\uppercase\expandafter{%
\romannumeral #1}}
\begin{document}

\title[Bosonic representation of spin operators]
{Bosonic representation of spin operators in the field-induced
critical phase of spin-1 Haldane chains}

\author{Masahiro Sato}

\address{Synchrotron Radiation Research Center, Japan Atomic 
Energy Agency, Sayo, Hyogo 679-5148, Japan and CREST JST, Japan}
%\ead{m_sato@spring8.or.jp}
\begin{abstract}
A bosonized-field-theory representation of spin operators in the 
uniform-field-induced Tomonaga-Luttinger-liquid (TLL) phase 
in spin-1 Haldane chains is formulated by means of the non-Abelian 
(Tsvelik's Majorana fermion theory) and the Abelian bosonizations 
and Furusaki-Zhang technique [Phys. Rev. B {\bf 60}, 1175 (1999)]. 
It contains massive magnon fields as well as massless boson fields. 
From it, asymptotic forms of spin correlation functions in the 
TLL phase are completely determined. Applying the formula, 
we further discuss effects of perturbations 
(bond alternation, single-ion anisotropy terms, staggered fields, etc) 
for the TLL state, and the string order parameter.  
Throughout the paper, we often consider in some detail how the 
symmetry operations of the Haldane chains are translated in the 
low-energy effective theory. 
\end{abstract}

\pacs{75.10.Jm,75.10.Pq,75.50.Ee}
%75.10.Jm: Quantized spin models
%75.10.Pq: Spin chain models
%75.50.Ee: Antiferromagnetics
%75.40.Cx: Static properties (order parameter, static susceptibility,
%heat capacities, critical exponents, etc.)
%75.30.Kz: Magnetic phase boundaries (including magnetic transitions,
%metamagnetism, etc.)


\maketitle

%%%%%%%%%%%%%%%%%%%%%%%%%%%%%%%%%%%%%%%%%%%%%%%%%%%%%%%%
%%%%%%%%%%%%%%%%%%%%%%%%%%%%%%%%%%%%%%%%%%%%%%%%%%%%%%%%
%%%%%%%%%%%%%%%%%%%%%%%%%%%%%%%%%%%%%%%%%%%%%%%%%%%%%%%%
%%%%%%%%%%%%%%%%%%%%%%%%%%%%%%%%%%%%%%%%%%%%%%%%%%%%%%%%
%%%%%%%%%%%%%%%%%%%%%%%%%%%%%%%%%%%%%%%%%%%%%%%%%%%%%%%%
%%%%%%%%%%%%%%%%%%%%%%%%%%%%%%%%%%%%%%%%%%%%%%%%%%%%%%%%

\section{Introduction}\label{Int}
%Today, there are a number of (quasi) one-dimensional (1D) magnets. 
%Among them, spin-1/2 and spin-1 antiferromagnetic (AF) 
%chains~\cite{Aff-Lec,Mag} have played a significant role of the
%understanding and the development of low-dimensional magnetism.

For (quasi) one-dimensional (1D) magnets, 
sophisticated theoretical tools have been developed in the last
decades; for example, bosonizations~\cite{Gogo,Tsv,Gia}, 
conformal field theory (CFT)~\cite{CFT}, Bethe ansatz~\cite{QIS,Ta-solv}, 
and several numerical methods (exact diagonalization, 
density-matrix renormalization group (DMRG), quantum Monte
Carlo method (QMC), etc). Thanks to them, we have obtained a deep
understanding of 1D spin systems. 
%Particularly, the above strategies are
%powerful in the spin-1/2 1D magnets. 
Of course, they have succeeded in explaining a number of experimental 
results of quasi 1D magnets. 
In this paper, we try to further develop
%we try to contribute
%to the fruther development of a low-energy field theory description 
%for the spin-1 antiferromagnetic (AF) chain with a uniform field, 
a low-energy field theory description 
for spin-1 antiferromagnetic (AF) chains in a uniform magnetic field, 
which are a fundamental model in 1D spin systems.

Before discussing spin-1 AF chains, for comparison, let us briefly review 
the well-established low-energy effective theory (so-called Abelian 
bosonization) for the spin-1/2 XXZ chain~\cite{Aff-Lec,Mag}, 
which is one of the well-investigated and realistic models 
in spin systems. The Hamiltonian is defined by 
\begin{eqnarray}
\label{xxz}
\hat{\cal H}_{\rm xxz} &=& J\sum_j\left[\frac{1}{2}
(S_j^+S_{j+1}^-+{\rm h.c.})+\Delta_z S_j^zS_{j+1}^z\right]-H\sum_j S_j^z,
\end{eqnarray}
where $\vec S_j$ is spin-1/2 operator on site $j$, $J>0$ is the exchange
interaction, $\Delta_z$ is the anisotropy parameter ($\Delta_z=1$ case
is the SU(2) Heisenberg chain), and $H\geq 0$ is a uniform field.
In a wide parameter regime including the zero-field line 
($1<\Delta_z\leq 1$ and $H=0$)~\cite{Gia,Mag}, 
the low-energy physics is governed by a one-component 
Tomonaga-Luttinger liquid (TLL) phase, where the spin correlation
functions are of an algebraical decay type.  
For this critical phase, a beautiful field theory 
description has been established as follows. 
The low-energy effective Hamiltonian
is a massless Gaussian model, 
\begin{eqnarray}
\label{Gaussian}
\hat{\cal H}_{\rm xxz}^{\rm eff} &=& \int dx \frac{\bar v}{2}
\left[{\bar K}(\partial_x{\bar \theta})^2
+\frac{1}{\bar K}(\partial_x{\bar \phi})^2\right],
\end{eqnarray}
where $x=j\times a_0$ ($a_0$ is the lattice constant), 
$\bar\phi(x)$ is the scalar bosonic field, $\bar\theta(x)$ is the dual 
of $\bar\phi$,  
%(the equal-time commutation relation between $\bar\phi$ and
%$\bar\theta$ is 
%$[\bar\theta(x),\bar\phi(y)]=\frac{i}{2}{\rm sign}(y-x)$), 
$\bar v$ is equal to the massless spinon 
velocity of the chain~(\ref{xxz}), and $\bar K$ is the TLL parameter 
($\bar K=1$ and $\bar K=1/2$ respectively correspond to the XY (free fermion)
point $\Delta_z=0$ and the SU(2) Heisenberg one $\Delta_z=1$)~\cite{radius}. 
In this effective theory framework, the spin operator $\vec S_j$ is
approximated~\cite{Gia,Aff-Lec,Mag,H-F} as 
\begin{eqnarray}
\label{spin_xxz}
S_j^z/a_0 &\approx& \frac{\bar M}{a_0}+\frac{1}{\sqrt{\pi}}\partial_x{\bar\phi}
+(-1)^j A_1^z\sin\left(\sqrt{4\pi}{\bar\phi}
 + 2\pi\bar M\frac{x}{a_0}\right)+\cdots,\nonumber\\
S_j^+/a_0  &\approx& e^{i\sqrt{\pi}\bar\theta}\left[(-1)^j A_0
+A_1\sin\left(\sqrt{4\pi}{\bar\phi}
 + 2\pi\bar M\frac{x}{a_0}\right)+\cdots\right], 
\end{eqnarray}
where $\bar M$ is the magnetization per site $\langle S_j^z\rangle$, 
and $A_n$ and $A_n^z$ are nonuniversal real constants.
The magnetization $\bar M$, the velocity $\bar v$, and the TLL parameter
$\bar K$ can be determined as a function of the parameters 
$(J,\Delta_z,H)$, via the Bethe ansatz integral equations for the XXZ
chain. If these three
parameters are given, one can correctly know the asymptotic behavior 
of any spin correlation functions through the formula~(\ref{spin_xxz}). 
Since the accurate values of $A_n$ and $A_n^z$ 
have recently been estimated~\cite{Lu-Za,H-F}, 
it is also possible to obtain the amplitudes of
the spin correlation functions. This effective theory,
especially the bosonized spin formula~(\ref{spin_xxz}), is very useful
in analyzing not only the XXZ chain itself, but also several
perturbation effects (bond alternations, next-nearest-neighboring
interactions, anisotropy terms, etc) for the XXZ chain and coupled
spin-1/2 chain systems.

Now, we turn to spin-1 AF chains and their effective theories. 
In this paper, we mainly consider 
the following simple spin-1 AF Heisenberg chain, 
\begin{eqnarray}
\label{spin1AF}
\hat{\cal H}_{S=1} &=& J\sum_j\vec S_j\cdot\vec S_{j+1}-H\sum_j S_j^z,
\end{eqnarray}
where $\vec S_j$ is ``spin-1'' operator on site $j$.
In contrast to the
spin-1/2 case, the spin-1 AF chain~(\ref{spin1AF}) without a uniform field 
($H=0$) possesses a finite excitation gap (call Haldane gap) on its 
disordered ground state, 
which is described by the famous valence-bond-solid
picture~\cite{AKLT}. 
The low-lying excitation consists of a massive spin-1 magnon
triplet around wave number $p=\pi/a_0$. 
The spin correlation functions exhibit an exponential decay.  
This drastic difference between the spin-1/2 and spin-1 cases was
first predicted by Haldane in 1983~\cite{Hal}: he developed the nonlinear sigma
model (NLSM) approach~\cite{Frad,Auer,Aff-Lec} for spin systems in order to 
explain the above low-energy physics of the spin-1 chain.
After Haldane's conjecture, based on the non-Abelian bosonization
approach to 1D spin systems~\cite{Aff_NAB,Za-Fa,Aff-Hal}, 
Tsvelik proposed a Majorana (real) fermion
theory~\cite{Tsv_NAB,Gogo,Tsv,Ess-Aff}, which also has an ability 
to demonstrate the low-energy properties of the spin-1 AF chains. 
Both theories are now a standard field theory method 
of studying 1D spin-1 AF systems~\cite{note_sch}, 
like the Abelian bosonization for spin-1/2 chains.

When a uniform field $H$ is applied in the spin-1 SU(2) AF chain~(\ref{spin1AF}), 
the low-lying magnon bands are split due to the Zeeman coupling. 
Provided that the bottom of one band 
crosses the energy level of the disordered ground state, a magnon
condensed state and a finite magnetization
occur~\cite{Aff-bose,Aff-magBEC,Kon,Fath}. 
It is well known that this quantum phase transition is of a
commensurate-incommensurate (C-IC) type~\cite{Sch_comincom,Gia}, 
and the resultant phase can be regarded as a TLL.     
Therefore, it is expected that, just like the formula~(\ref{spin_xxz}), 
one can obtain a bosonized-spin expression in this TLL phase. 
However, as far as we know, 
any clear bosonization formulas of spin-1 operators in this
magnon-condensed phase have not been established yet. 
Namely, such a formula has not been argued enough so far~\cite{note1}.  
%Namely, compared to the case of spin-1/2 chains, 
%the effective field theory for spin-1 AF chains has been less completed.
%, at least in our opinion. 

The main purpose of this paper is to derive a satisfactory bosonization 
formula for spin-1 operators in the above TLL phase. 
To this end, we start with Tsvelik's Majorana fermion theory. This theory 
makes it possible to treat the Zeeman term in a nonperturbative way. 
Moreover, it can carefully deal with the uniform component of spins 
(the part around wave number $p\sim 0$) as well as the staggered
ones, in contrast to the NLSM approach.  
(We explain these properties of the fermion theory in later sections). 
To make the new bosonization formula, 
the bosonized-spin expression for the uniform-field-induced TLL phase 
in a two-leg spin-1/2 AF ladder~\cite{Shelton} and its derivation, which
was performed by Furusaki and Zhang~\cite{Fu-Zh}, 
is very instructive and helpful. We will rely on these results. 
%in the construction of the
%bosonization formula for the spin-1 operator. 
We will often discuss the symmetry correspondences between the original
spin-1 AF chain~(\ref{spin1AF}) and the effective field theory. 
Like the formula~(\ref{spin_xxz}), 
our resulting formula would be powerful in studying
various (quasi) 1D spin-1 AF systems with magnons condensed. 
Recently, magnon-condensed states in quasi 1D gapped magnets, 
including spin-1 compounds such as 
NDMAP~\cite{Chen,Zhel1,Zhel2,Hagi,Zhel3,Tsuji}, NDMAZ~\cite{Zhel4},
TMNIN~\cite{Goto} and NTENP~\cite{Hagi1,Hagi2}, 
have been intensively investigated by some experiment
groups. Our new formula hence is expected to be effective for 
understanding results of such experiments.

The organization of the rest of the paper is as follows. 
In Sec.~\ref{fermion_review}, we review Tsvelik's Majorana fermion
theory for the spin-1 AF chain without external fields. 
Section~\ref{inducedTLL} is devoted
to discussing the effective field theory for the uniform-field-induced 
TLL in spin-1 AF chains, which is based on the Majorana fermion theory. 
These two sections are needed as the preparation to construct a field-theory
formula for spin operators in the TLL phase, 
although several parts of them have been already 
discussed in literature~\cite{Gogo,Tsv,Tsv_NAB,Ess-Aff,MS05_1}.    
In Sec.~\ref{spin_op}, which is the main content of this paper, 
employing the contents of Secs.~\ref{fermion_review} and 
\ref{inducedTLL}, we derive the formula of spin operators. 
%Furthremore, using it, we consider the symmetires of the
%spin-1 chains. 
Moreover, using it, we estimate the asymptotic forms of
spin correlation functions in the TLL phase. 
%Throughout these three sections,
%we emphasize how symmetries of the spin-1 AF chain~(\ref{spin1AF}) are
%represented in the effective field theory framework. 
Section~\ref{Application} provides easy applications of our 
field-theory formula of spin operators. We consider the nonlocal 
string order parameter~\cite{Naka,MS05_1}, %Nijs,Hatsu,Kenn,
effects of some perturbation terms 
(bond alternation, a few anisotropy terms, etc) 
for the TLL state, and the magnon-decay process caused by anisotropic
perturbations.  
We summarize the results of this paper in Sec.~\ref{Summary}. 
Appendices may prove useful when reading the main text.

%%%%%%%%%%%%%%%%%%%%%%%%%%%%%%%%%%%%%%%%%%%%%%%%%%%%%%%%
%%%%%%%%%%%%%%%%%%%%%%%%%%%%%%%%%%%%%%%%%%%%%%%%%%%%%%%%
%%%%%%%%%%%%%%%%%%%%%%%%%%%%%%%%%%%%%%%%%%%%%%%%%%%%%%%%
%%%%%%%%%%%%%%%%%%%%%%%%%%%%%%%%%%%%%%%%%%%%%%%%%%%%%%%%
%%%%%%%%%%%%%%%%%%%%%%%%%%%%%%%%%%%%%%%%%%%%%%%%%%%%%%%%
%%%%%%%%%%%%%%%%%%%%%%%%%%%%%%%%%%%%%%%%%%%%%%%%%%%%%%%%
%%%%%%%%%%%%%%%%%%%%%%%%%%%%%%%%%%%%%%%%%%%%%%%%%%%%%%%%
%%%%%%%%%%%%%%%%%%%%%%%%%%%%%%%%%%%%%%%%%%%%%%%%%%%%%%%%
%%%%%%%%%%%%%%%%%%%%%%%%%%%%%%%%%%%%%%%%%%%%%%%%%%%%%%%%
%%%%%%%%%%%%%%%%%%%%%%%%%%%%%%%%%%%%%%%%%%%%%%%%%%%%%%%%

\section{Spin-1 AF chain without external fields}
\label{fermion_review}
In this section, we review the non-Abelian bosonization (Tsvelik's
Majorana fermion theory) method for the spin-1 Heisenberg AF chain with
$H=0$, especially focusing on the symmetries of the chain.

\subsection{WZNW model and Majorana fermion theory}
\label{WZNW_fermion}
First, we define the spin-1 bilinear-biquadratic chain, 
\begin{eqnarray}
\label{bilbiq}
\hat{\cal H}_{\beta} &=& J\sum_j\left[\vec S_j\cdot\vec S_{j+1}
+\beta(\vec S_j\cdot\vec S_{j+1})^2\right].
\end{eqnarray}
Several studies~\cite{S1bilbiq} reveal that the system 
belongs to the ``Haldane phase'' in the regime $|\beta|< 1$, 
%which contains the spin-1 Heisenberg chain ($\delta=0$), 
and the points $\beta=\pm 1$ correspond to a quantum criticality. In the
Haldane phase, the ground state is disordered 
(namely, all symmetries are conserved) and the lowest excitations have 
a finite Haldane gap on it. It is shown~\cite{Aff_NAB,Avd,Tak-Bab}
that the low-energy physics of the critical point $\beta=-1$ is 
captured by an SU(2) level-2 Wess-Zumino-Novikov-Witten (WZNW) 
model~\cite{Gogo,Tsv}, a model of CFT with central charge $c=3/2$, 
plus irrelevant perturbations preserving the spin SU(2) symmetry. 
This WZNW model possesses a Majorana
(real) fermion description, i.e., the WZNW model with $c=3/2$ is equivalent to
three copies of massless real fermions, each of which is regarded as a
critical Ising system (a minimal model of CFT) with 
$c=1/2$~\cite{Za-Fa,Shelton,Gogo,Tsv}.
Using the fermion (or Ising) picture, 
one can represent the Hamiltonian of the WZNW model as follows:
\begin{eqnarray}
\label{WZNW_fermion}
\hat{\cal H}_{\rm wznw} &=& \int dx \,\frac{i}{2}v_0\sum_{q=1,2,3}
\Big[\xi_L^q\partial_x\xi_L^q-\xi_R^q\partial_x\xi_R^q\Big],
\end{eqnarray}
where $\xi_{L(R)}^q(x)={\xi_{L(R)}^q}^\dag(x)$ is the left (right)
moving field of the $q$th real fermion 
with scaling dimension $\Delta_s=1/2$, and 
$v_0$ is the ``light'' velocity of the fermions. We here normalize the 
fermions $\xi_{L(R)}^q$ through the anticommutation relation 
$\{\xi_\nu^q(x),\xi_{\nu'}^{q'}(y)\}=\delta_{\nu,\nu'}\delta_{q,q'}\delta(x-y)$, 
where the delta function stands for $\delta_{x,y}/\alpha$ 
[$\alpha$ is a constant of ${\cal O}(a_0)$]. 
At the WZNW point $\beta=-1$, the spin operator is approximated as the
following sum of the uniform and the staggered
components~\cite{Aff_NAB,Tsv_NAB,Gogo,Tsv},  
\begin{eqnarray}
\label{spin_fermion}
S_j^q/a_0 \approx J^q(x) + (-1)^j C_0 N^q(x), \,\,\,\,
(q=1,2,3\,\, {\rm or}\,\, x,y,z),
\end{eqnarray}
where
\begin{eqnarray}
J^q(x)=J_L^q(x)+J_R^q(x),\hspace{0.5cm} 
J_{L(R)}^q=-i\xi_{L(R)}^{q+1}\xi_{L(R)}^{q+2}, \,\,\,(q+3=q),\nonumber\\
\left\{
\begin{array}{ccc}
N^1=\mu^1\sigma^2\mu^3 \\
N^2=\sigma^1\mu^2\mu^3 \\
N^3=\mu^1\mu^2\sigma^3
\end{array}
\right.,
\end{eqnarray}
and $C_0$ is a nonuniversal constant. The quantity $J_{L(R)}^q$ is 
equivalent to the left (right) component of the SU(2) current in 
the WZNW model with $\Delta_s=1$, and  
$\sigma^q$ ($\mu^q$) is the $q$th-Ising order (disorder)
field with $\Delta_s=2/16$. 
%These facts are the starting point of the Tsvelik's theory. 

%The spin-1 SU(2) Heisenberg chain~(\ref{spin1AF}), our target, 
%lies in the point $\delta=0$. 
According to Tsvelik~\cite{Tsv_NAB,Aff-Hal,Gogo,Tsv}, transferring
from the WZNW point $\beta=-1$ to the Heisenberg one $\beta=0$ 
corresponds to adding two SU(2)-symmetric (see the next subsection) 
perturbation terms, a fermion mass term and a marginally irrelevant 
current-current interaction, 
to the WZNW model~(\ref{WZNW_fermion}). 
Namely, he proposed that the low-energy and long-distance 
properties of the spin-1 Heisenberg chain~(\ref{spin1AF}) 
are govern by the effective Hamiltonian, 
\begin{eqnarray}
\label{S1Hisenberg_eff}
\hat{\cal H}_{S=1}^{\rm eff} &=& \int dx \sum_{q}
\left[\frac{i}{2}v_0\left(\xi_L^q\partial_x\xi_L^q-\xi_R^q\partial_x\xi_R^q\right)
+mi\xi_L^q\xi_R^q  -\lambda J_L^qJ_R^q\right],
\end{eqnarray}
where we may set $m>0$ and $\lambda>0$, which 
guarantees the irrelevancy of the $\lambda$ term.
From the viewpoint of the spin-1 AF chain, 
fermions $\xi_{L,R}^q$ stand for the low-lying magnon
triplet around wave number $p=\pi/a_0$, the mass parameter $m$
corresponds to the Haldane gap, and the $\lambda$ term represents 
the inter-magnon interaction. 
Renormalizing the $\lambda$ interaction
effect into the kinetic and the mass terms, or simply neglecting the 
$\lambda$ term, one can determine the values of $v_0$ and $m$ from the
numerically-estimated magnon dispersion of the spin-1 Heisenberg chain; 
such a procedure provides $v_0\approx 2.49 Ja_0$ and 
$m\approx 0.41 J$~\cite{Qin,To-Ka}.  
One should interpret that the inequality $m>0$ corresponds to the
disordered phase in the Ising-model picture, in which 
$\langle\mu^q\rangle\neq 0$ and $\langle\sigma^q\rangle= 0$. 
For the disordered phase, Ising-field correlation
functions~\cite{Wu,Shelton,Tsv_NAB} are shown to be  
\begin{eqnarray}
\label{Ising_corr} 
\langle\sigma^q(x)\sigma^q(0)\rangle &\approx& 
\frac{B_1}{\sqrt{2\pi|x|/\xi_c}}
\left(1+{\cal O}(|x/\xi_c|^{-1})\right)e^{-|x|/\xi_c}
+{\cal O}(e^{-3|x|/\xi_c}),\nonumber\\
\langle\mu^q(x)\mu^q(0)\rangle &\approx& 
B_1\left[1-\left(\frac{1}{16\pi|x|/\xi_c}+{\cal
	    O}(|x/\xi_c|^{-2})\right)e^{-2|x|/\xi_c}\right]\nonumber\\
&&+{\cal O}(e^{-4|x|/\xi_c}),
\end{eqnarray}
at a long distance, $x\gg \xi_c=v_0/m$~\cite{note2}. 
Here, $B_1$ is a nonuniversal constant, and $\xi_c\approx 6 a_0$ 
is the correlation length. 
If we assume that the field-theory formula~(\ref{spin_fermion}) is valid
even at the Heisenberg point $\beta=0$, that formula 
and the result~(\ref{Ising_corr}) lead to 
the following asymptotic behavior of the spin correlation 
functions of the spin-1 Heisenberg chain:
\begin{eqnarray}
\label{spin1_corr} 
\langle S_j^qS_0^q\rangle &\approx& 
(-1)^j\frac{B_2}{\sqrt{|x|/\xi_c}}e^{-|x|/\xi_c}
+\frac{B_3}{|x|/\xi_c}e^{-2|x|/\xi_c}+\cdots,
\end{eqnarray}
where we used Eqs.~(\ref{xi_2point}) and (\ref{xi_2point_part2}) to 
derive the second term, and $B_{2,3}$ are nonuniversal constants.
The exponential parts in Eq.~(\ref{spin1_corr}) are consistent 
with those predicted by the NLSM approach~\cite{Aff-So}, but the
prefactor ($\propto 1/x$) of the second uniform term differs from that
of the NLSM, which predicts a prefactor proportional to $1/x^2$.     
From the calculation of DMRG~\cite{Aff-So}, we may conclude 
$B_2\approx 0.25$ and $B_3\ll B_2$.

%%%%%%%%%%%%%%%%%%%%%%%%%%%%%%%%%%%%%%%%%%%%%%%%%%
%%%%%%%%%%%%%%%%%%%%%%%%%%%%%%%%%%%%%%%%%%%%%%%%%%
%%%%%%%%%%%%%%%%%%%%%%%%%%%%%%%%%%%%%%%%%%%%%%%%%%
%%%%%%%%%%%%%%%%%%%%%%%%%%%%%%%%%%%%%%%%%%%%%%%%%%

\subsection{Symmetries of the spin-1 AF Heisenberg chain}
\label{Sym_s1}
The effective Hamiltonian~(\ref{S1Hisenberg_eff}) and the field-theory
expression of the spin~(\ref{spin_fermion}) tell us how the symmetries of
the spin-1 Heisenberg chain are translated in the field theory
language~\cite{Aff_NAB,Aff-Hal,Ess-Aff}. 
Here, we consider four kinds of symmetry: the spin rotation 
$\vec S_j\to {\cal R}\vec S_j$ ($\cal R$ is an SO(3) matrix), the 
one-site translation $\vec S_j\to \vec S_{j+1}$, 
the time reversal $\vec S_j\to -\vec S_j$, and 
the site-parity transformation $\vec S_j\to \vec S_{-j}$. Note that the
link-parity symmetry is the combination of the one-site translational 
and site-parity symmetries.

(\rnum{1}) The spin rotation may be realized by 
\begin{eqnarray}
\label{rotation} 
\vec \xi_{L,R}\to{\cal R} \vec\xi_{L,R}, \hspace{1cm} 
\vec N\to {\cal R}\vec N,
\end{eqnarray}
where $\vec \xi_\nu={}^t(\xi_\nu^1,\xi_\nu^2,\xi_\nu^3)$ 
and $\vec N={}^t(N^1,N^2,N^3)$. The ``rotation'' of fermions
induces $\vec J_\nu\to{\cal R}\vec J_\nu$ and $\vec J\to{\cal R}\vec J$, where 
$\vec J_\nu={}^t(J_\nu^1,J_\nu^2,J_\nu^3)$ and $\vec J={}^t(J^1,J^2,J^3)$. 
The latter transformation in Eq.~(\ref{rotation}) could be
regarded as 
\begin{eqnarray}
\label{rotation_sigma} 
\vec \sigma\to{\cal R} \vec\sigma,
\end{eqnarray}
where $\vec \sigma={}^t(\sigma^2,\sigma^1,\sigma^3)$. 
If both the mass and the interaction terms are absent (i.e., the system
is completely critical), 
the chiral SO(3)$\times$SO(3) symmetry (independent rotations of the 
left and right fields) appears.

(\rnum{2}) The one-site translation may correspond to 
\begin{eqnarray}
\label{one-site} 
\vec \xi_{L,R}(x)\to - \vec\xi_{L,R}(x+a_0),\hspace{1cm} 
\vec N(x)\to - \vec N(x+a_0).
\end{eqnarray}
Since the fermions $\xi_{L,R}^q$ describe the magnon excitations around
wave number $p=\pi/a_0$ in the original chain~(\ref{spin1AF}), 
we should insert $-1$ in the first transformation of Eq.~(\ref{one-site}). 
The second transformation can be realized by 
\begin{eqnarray}
\label{one-site-staggered} 
\left\{
\begin{array}{cc}
\vec \sigma(x)\to - \vec \sigma(x+a_0)\\
\vec \mu(x)\to  \vec \mu(x+a_0)
\end{array}
\right.,
\hspace{0.5cm}{\rm or}\hspace{0.5cm}
\left\{
\begin{array}{cc}
\vec \sigma(x)\to - \vec \sigma(x+a_0)\\
\vec \mu(x)\to - \vec \mu(x+a_0)
\end{array}
\right.,
\end{eqnarray}
where $\vec\mu={}^t(\mu^1,\mu^2,\mu^3)$.
Due to this transformation, 
the factor $(-1)^j$ in front of the staggered component of the
spin~(\ref{spin_fermion}) changes to $(-1)^{j+1}$.

(\rnum{3}) The time reversal may correspond to
\begin{eqnarray}
\label{time} 
\vec \xi_{L,R} \to \vec\xi_{R,L},
\hspace{1cm} 
\vec N\to - \vec N, \hspace{1cm} i\to -i.
\end{eqnarray}
The second mapping $\vec N\to - \vec N$ could be interpreted as 
\begin{eqnarray}
\label{parity-staggered} 
\left\{
\begin{array}{cc}
\vec \sigma\to - \vec \sigma\\
\vec \mu\to  \vec \mu
\end{array}
\right.,
\hspace{0.5cm}{\rm or}\hspace{0.5cm}
\left\{
\begin{array}{cc}
\vec \sigma\to - \vec \sigma\\
\vec \mu\to - \vec \mu
\end{array}
\right..
\end{eqnarray}

(\rnum{4}) As a proper mapping of the continuous fields 
towards the site-parity transformation, we can propose  
\begin{eqnarray}
\label{parity}
\left\{
\begin{array}{ccc}
\vec \xi_L(x)\to \mp\vec\xi_R(-x)\\
\vec \xi_R(x)\to \pm\vec\xi_L(-x)
\end{array}
\right.,  
\hspace{1cm} 
\vec N(x)\to \vec N(-x).
\end{eqnarray}

All the transformations (\rnum{1})-(\rnum{4}) leaves 
the effective Hamiltonian~(\ref{S1Hisenberg_eff}) invariant. 
In other words, 
the symmetries (\rnum{1})-(\rnum{4}) allow the existence of 
the mass term and the current-current interaction one.

%%%%%%%%%%%%%%%%%%%%%%%%%%%%%%%%%%%%%%%%%%%%%%%%%%%%%%%%
%%%%%%%%%%%%%%%%%%%%%%%%%%%%%%%%%%%%%%%%%%%%%%%%%%%%%%%%
%%%%%%%%%%%%%%%%%%%%%%%%%%%%%%%%%%%%%%%%%%%%%%%%%%%%%%%%
%%%%%%%%%%%%%%%%%%%%%%%%%%%%%%%%%%%%%%%%%%%%%%%%%%%%%%%%
%%%%%%%%%%%%%%%%%%%%%%%%%%%%%%%%%%%%%%%%%%%%%%%%%%%%%%%%

\subsection{Dirac fermion and Abelian bosonization}
\label{2RealFermions_bosonization}
Besides the relationship between the WZNW model 
and three copies of massless real fermions, it is well known that 
two copies of the real fermions are equivalent to a 
massless Gaussian theory with TLL parameter $1$ or 
a massless Dirac (complex) fermion, 
except for a few subtle aspects~\cite{CFT}. Namely, one can bosonize 
two copies of the real fermions. 
In this subsection, we apply such a bosonization procedure 
for the effective theory~(\ref{S1Hisenberg_eff}). 
The results are useful in the later sections.

Let us define the following Dirac fermion,
\begin{eqnarray}
\label{Dirac}
\psi_L=\frac{1}{\sqrt{2}}(\xi_L^1+i\xi_L^2),\hspace{1cm}
\psi_R=\frac{1}{\sqrt{2}}(\xi_R^1+i\xi_R^2).
\end{eqnarray}
which obey $\{\psi_\nu(x),\psi_{\nu'}(y)\}=0$ and 
$\{\psi_\nu(x),\psi_{\nu'}^\dag(y)\}=\delta_{\nu,\nu'}\delta(x-y)$. 
Using this, we can rewrite the bilinear (free) part of the 
Hamiltonian~(\ref{S1Hisenberg_eff}) as follows:  
\begin{eqnarray}
\label{Dirac_free}
\hat{\cal H}_{\rm free} &=& \int \,dx\, 
iv_0\left(\psi_L^\dag\partial_x\psi_L-\psi_R^\dag\partial_x\psi_R\right)
+mi(\psi_L^\dag\psi_R-\psi_R^\dag\psi_L)\nonumber\\
&&+\frac{i}{2}v_0\left(\xi_L^3\partial_x\xi_L^3-\xi_R^3\partial_x\xi_R^3\right)
+mi\xi_L^3\xi_R^3\nonumber\\
&=&\hat{\cal H}[\psi]+\hat{\cal H}[\xi^3]. 
\end{eqnarray}
Applying the standard Abelian bosonization
method (see \ref{Abelian}) 
to the Dirac-fermion part $\hat{\cal H}[\psi]$, we obtain the following 
sine-Gordon Hamiltonian, 
\begin{eqnarray}
\label{Dirac_to_Boson}
\hat{\cal H}[\psi] &\to& \hat{\cal H}[\Phi]\,=\, \int \,dx\, 
\frac{v_0}{2}\left[(\partial_x\Theta)^2+(\partial_x\Phi)^2\right]
+\frac{m}{\pi\alpha}\cos(\sqrt{4\pi}\Phi),
\end{eqnarray}
where $\Phi$ is the scalar field, $\Theta$ is the dual of $\Phi$,
$\alpha\sim a_0$ is the nonuniversal short-distance cut off, 
the relevant cos term bears the Haldane gap, and 
we simply dropped the Klein factors. The fields 
$(\xi_\nu^1,\xi_\nu^2)$ and $(\sigma^{1,2},\mu^{1,2})$
in the spin operator~(\ref{spin_fermion}) 
can also be rewritten in the Dirac-fermion or the boson languages.
The uniform component in Eq.~(\ref{spin_fermion}) is  
\begin{eqnarray}
\label{spin-uniform}
J_L^1+J_R^1 &=& \frac{1}{\sqrt{2}}(\psi_L^\dag-\psi_L)\xi_L^3
+\frac{1}{\sqrt{2}}(\psi_R^\dag-\psi_R)\xi_R^3\nonumber\\
&=& \frac{i}{\sqrt{\pi\alpha}}
\left[\zeta_L\xi_L^3\sin(\sqrt{\pi}(\Phi+\Theta))
-\zeta_R\xi_R^3\sin(\sqrt{\pi}(\Phi-\Theta))\right],
\nonumber\\
J_L^2+J_R^2 &=& \frac{i}{\sqrt{2}}(\psi_L^\dag+\psi_L)\xi_L^3
+\frac{i}{\sqrt{2}}(\psi_R^\dag+\psi_R)\xi_R^3\nonumber\\
&=& \frac{i}{\sqrt{\pi\alpha}}
\left[\zeta_L\xi_L^3\cos(\sqrt{\pi}(\Phi+\Theta))
+\zeta_R\xi_R^3\cos(\sqrt{\pi}(\Phi-\Theta))\right],
\nonumber\\
J_L^3+J_R^3 &=& -:\psi_L^\dag\psi_L :- :\psi_R^\dag\psi_R: 
= -\frac{1}{\sqrt{\pi}}\partial_x\Phi,
\end{eqnarray}
where $\zeta_\nu$ is the Klein factor for the fermion $\psi_\nu$, 
and we assumed $\{\zeta_\nu,\xi_{\nu'}^3\}=0$. The symbol $:\,\,:$
denotes a normal-ordered product; 
$:\psi_\nu^\dag\psi_\nu:=\psi_\nu^\dag\psi_\nu
-\langle\psi_\nu^\dag\psi_\nu\rangle$.
From the formula~(\ref{boso_3}), the staggered component of 
the spin~\cite{note_compact} is bosonized as 
\begin{eqnarray}
\label{spin-stagg}
N^1 &=& \mu^3\sin(\sqrt{\pi}\Theta),\nonumber\\
N^2 &=& \mu^3\cos(\sqrt{\pi}\Theta),\nonumber\\
N^3 &=& \sigma^3\cos(\sqrt{\pi}\Phi).
\end{eqnarray}

Employing Eqs.~(\ref{Dirac})-(\ref{spin-stagg}) 
and Abelian bosonization formulas (see~\ref{Abelian}), 
let us consider how symmetry operations of the original 
spin-1 chain~(\ref{spin1AF}) are represented 
within the above Dirac-fermion and boson framework, 
as in the last subsection.

(\rnum{1}) Let us define the following 
U(1) rotation around the spin z axis,   
\begin{eqnarray}
\label{U(1)rotation}
\left(
\begin{array}{cc}
S_j^x\\
S_j^y\\
S_j^z
\end{array}
\right)
&\to&
\left(
\begin{array}{ccc}
\cos\gamma & -\sin\gamma & 0 \\
\sin\gamma & \cos\gamma  & 0 \\
0& 0& 1
\end{array}
\right)
\left(
\begin{array}{cc}
S_j^x\\
S_j^y\\
S_j^z
\end{array}
\right),
\end{eqnarray}
where $\gamma$ is a real number.
One can easily find that this corresponds to 
\begin{eqnarray}
\label{U(1)_Dirac_boson}
\psi_{L,R}\to e^{i\gamma}\psi_{L,R}, \hspace{1cm}
\Theta(x)\to \Theta(x)-\frac{\gamma}{\sqrt{\pi}}.
\end{eqnarray}
This transformation is nothing but the U(1) symmetry that is explicitly 
preserved via the Abelian bosonization procedure.

(\rnum{2}) The one-site translation $\vec S_j\to\vec S_{j+1}$ 
could be realized as 
\begin{eqnarray}
\label{one-site_Dirac_boson}
\left\{
\begin{array}{cc}
\psi_{L,R}(x)\to-\psi_{L,R}(x+a_0)\\
\xi_{L,R}^3(x)\to- \xi_{L,R}^3(x+a_0)
\end{array}
\right.,\hspace{0.5cm}
\left\{
\begin{array}{ccc}
\Theta(x)\to \Theta(x+a_0)+\sqrt{\pi}\\
\Phi(x) \to  \Phi(x+a_0)
\end{array}
\right.,
\nonumber\\
\left\{
\begin{array}{ccc}
\sigma^3(x)  \to  -\sigma^3(x+a_0)\\
\mu^3(x)  \to  \mu^3(x+a_0)
\end{array}
\right..
\end{eqnarray}
The second and third transformations change the staggered factor $(-1)^j$ 
in Eq.~(\ref{spin_fermion}) into $(-1)^{j+1}$.
These two correspond to the first proposal in Eq.~(\ref{one-site-staggered}).

(\rnum{3}) The time reversal $\vec S_j\to-\vec S_j$ could be 
interpreted as
\begin{eqnarray}
\label{time_Dirac_boson}
\left\{
\begin{array}{ccc}
\psi_{L,R}\leftrightarrow\psi_{R,L}^\dag\\
\xi_{L,R}^3\to\xi_{R,L}^3\\
\end{array}
\right.,
\hspace{0.5cm}
\left\{
\begin{array}{ccc}
\zeta_{L,R}\to\zeta_{R,L}\\
\Phi_{L,R}\to-\Phi_{R,L}\\
\Phi\to-\Phi
\end{array}
\right.,\nonumber\\
\left\{
\begin{array}{ccc}
\sigma^3\to-\sigma^3\\
\mu^3\to-\mu^3
\end{array}
\right.,\hspace{0.5cm}i\to-i.
\end{eqnarray}
The second and third mappings are equivalent to the second proposal in
Eq.~(\ref{parity-staggered}).

(\rnum{4}) The site-parity transformation $\vec S_j\to\vec S_{-j}$ might
correspond to~\cite{note_Klein} 
\begin{eqnarray}
\label{parity_Dirac_boson}
\left\{
\begin{array}{ccc}
\psi_L(x)\to\mp\psi_R(-x)\\
\psi_R(x)\to\pm\psi_L(-x)\\
\xi_L^3(x)\to\mp\xi_R^3(-x)\\
\xi_R^3(x)\to\pm\xi_L^3(-x)
\end{array}
\right.,\hspace{0.5cm}
\left\{
\begin{array}{ccc}
\zeta_L\to\mp\zeta_R\\
\zeta_R\to\pm\zeta_L\\
\Phi(x)\to-\Phi(-x)\\
\Theta(x)\to\Theta(-x)
\end{array}
\right.,\nonumber\\
\left\{
\begin{array}{ccc}
\sigma^3(x)\to\sigma^3(-x)\\
\mu^3(x)\to\mu^3(-x)
\end{array}
\right..
\end{eqnarray}
%although the property of the Klein factors in
%Eq.~(\ref{parity_Dirac_boson}) seems to be strange. 

%%%%%%%%%%%%%%%%%%%%%%%%%%%%%%%%%%%%%%%%%%%%%%%%%%%%%%%%
%%%%%%%%%%%%%%%%%%%%%%%%%%%%%%%%%%%%%%%%%%%%%%%%%%%%%%%%
%%%%%%%%%%%%%%%%%%%%%%%%%%%%%%%%%%%%%%%%%%%%%%%%%%%%%%%%
%%%%%%%%%%%%%%%%%%%%%%%%%%%%%%%%%%%%%%%%%%%%%%%%%%%%%%%%
%%%%%%%%%%%%%%%%%%%%%%%%%%%%%%%%%%%%%%%%%%%%%%%%%%%%%%%%
%%%%%%%%%%%%%%%%%%%%%%%%%%%%%%%%%%%%%%%%%%%%%%%%%%%%%%%%
%%%%%%%%%%%%%%%%%%%%%%%%%%%%%%%%%%%%%%%%%%%%%%%%%%%%%%%%
%%%%%%%%%%%%%%%%%%%%%%%%%%%%%%%%%%%%%%%%%%%%%%%%%%%%%%%%
%%%%%%%%%%%%%%%%%%%%%%%%%%%%%%%%%%%%%%%%%%%%%%%%%%%%%%%%

\section{Effective field theory for uniform-field-induced 
TLL in spin-1 AF chain}
\label{inducedTLL}
Based on the Majorana fermion theory explained in the last section, 
we consider the uniform-field effects in the spin-1 AF
chain~(\ref{spin1AF}), especially the uniform-field-induced critical
phase. 
%discuss the uniform-field-induced critical phase in the spin-1 AF
%chain~(\ref{spin1AF}) within the field theory description. 

From the formula~(\ref{spin_fermion}), 
the Zeeman term in the chain~(\ref{spin1AF}) is approximated as 
the following fermion bilinear form,
\begin{eqnarray}
\label{Zeeman_fermion}
-H\sum_j S_j^z &\approx&\int\,dx\,\,iH(\xi_L^1\xi_L^2+\xi_R^1\xi_R^2).
\end{eqnarray}
Therefore, it is possible to nonperturbatively treat the Zeeman term,
together with the free part (\ref{Dirac_free}), within the field theory
scheme. This is a significant advantage of the Majorana fermion theory.  
Take notice here that the Zeeman coupling reduces the spin SU(2)
symmetry to the U(1) one around the spin z axis, 
and destroys the time-reversal symmetry; 
the remaining symmetries are the U(1), the one-site translation, 
and the site-parity ones. The right-hand side in
Eq.~(\ref{Zeeman_fermion}) is indeed not invariant under the 
time reversal and any spin rotations except for the U(1) one.

Supposing the inter-magnon interaction may be negligible 
at the starting point, the effective Hamiltonian 
for the spin-1 AF chain~(\ref{spin1AF}) with a finite $H$ is 
\begin{eqnarray}
\label{eff_finiteH}
\hat{\cal H}_{H}^{\rm eff} &=& \int\, dx\,
\Big[iv_0\left(\psi_L^\dag\partial_x\psi_L-\psi_R^\dag\partial_x\psi_R\right)
\nonumber\\
&&+mi(\psi_L^\dag\psi_R-\psi_R^\dag\psi_L)+H(\psi_L^\dag\psi_L+\psi_R^\dag\psi_R)
\nonumber\\
&&+\frac{i}{2}v_0\left(\xi_L^3\partial_x\xi_L^3-\xi_R^3\partial_x\xi_R^3\right)
+mi\xi_L^3\xi_R^3\Big].
\end{eqnarray}
In order to diagonalize this Hamiltonian, and see its band structure, 
we introduce the Fourier transformations of fermion fields as follows: 
\begin{eqnarray}
\label{Fourier_fermion}
\xi_\nu^3(x)=\frac{1}{\sqrt{L}}\sum_k e^{ikx}\tilde\xi_\nu^3(k),
\,\,\,&&\tilde\xi_\nu^3(k)=\frac{1}{\sqrt{L}}\int dx
\,e^{-ikx}\xi_\nu^3(x),\nonumber\\
\psi_\nu(x)=\frac{1}{\sqrt{L}}\sum_k e^{ikx}\tilde\psi_\nu(k),
\,\,\,&&\tilde\psi_\nu(k)=\frac{1}{\sqrt{L}}\int dx
\,e^{-ikx}\psi_\nu(x),
\end{eqnarray}
where $L=Na_0$ is the system size ($N$ : total number of site), 
and $k$ ($|k|<\Lambda$ : $\Lambda\sim a_0^{-1}$ is the 
ultraviolet cut off) is the
wave number: a related quantity $p=k+\pi/a_0$ is interpreted as 
the wave number of the original lattice system~(\ref{spin1AF}).
Since $\xi_\nu^3$ is real, $\tilde \xi_\nu^3{}^\dag(k)=\tilde\xi_\nu^3(-k)$
holds. New fermions $\tilde\xi_\nu^3$ and $\tilde\psi_\nu$ obey the
anticommutation relations 
$\{\tilde\xi_\nu^3(k),\tilde\xi_{\nu'}^3(k')\}=\delta_{\nu,\nu'}\delta_{k,-k'}$, 
$\{\tilde\psi_\nu(k),\tilde\psi_{\nu'}^\dag(k')\}=\delta_{\nu,\nu'}\delta_{k,k'}$,
and $\{\tilde\psi_\nu(k),\tilde\psi_{\nu'}(k')\}=0$. Substituting
Eq.~(\ref{Fourier_fermion}) into the Hamiltonian~(\ref{eff_finiteH}), 
we obtain
\begin{eqnarray}
\label{eff_k_space}
\hat{\cal H}_{H}^{\rm eff} &=&\sum_{|k|<\Lambda}
\tilde\Psi_k^\dag {\cal M}_k^\psi \tilde\Psi_k
+\sum_{0<k<\Lambda}\tilde\Xi_k^\dag {\cal M}_k^\xi\tilde\Xi_k, 
\end{eqnarray}
where $\tilde\Psi_k={}^t(\tilde\psi_R,\tilde\psi_L)$,
$\tilde\Xi_k={}^t(\tilde\xi_R^3,\tilde\xi_L^3)$, and 
\begin{eqnarray}
\label{matrix_k_space}
{\cal M}_k^\psi=\left(
\begin{array}{cc}
kv_0 + H & -im\\
im   & -kv_0 + H 
\end{array}
\right),&&\,\,\,\,\,\,
{\cal M}_k^\xi=\left(
\begin{array}{cc}
kv_0   & -im\\
im   & -kv_0 
\end{array}
\right).
\end{eqnarray}
Subsequently, performing the following Bogoliubov transformation, 
\begin{eqnarray}%Fourier_fermion,Bogoliu,Bogo_matrix
\label{Bogoliu}
\tilde\Psi_k= {\cal U}_k
\left(
\begin{array}{cc}
\tilde\eta_-(k)\\
\tilde\eta_+(k)
\end{array}
\right),&&\,\,\,\,\,\,
\tilde\Xi_k= {\cal U}_k
\left(
\begin{array}{cc}
\tilde\eta_0(k)\\
\tilde\eta_0(-k)
\end{array}
\right),
\end{eqnarray}
where 
\begin{eqnarray}
\label{Bogo_matrix}
{\cal U}_k=
\frac{m}{\sqrt{2\epsilon_0(k)}}\left(
\begin{array}{cc}
(\epsilon_0(k) - kv_0)^{-1/2}   & (\epsilon_0(k) + kv_0)^{-1/2}\\
i(\epsilon_0(k) - kv_0)^{1/2}/m & -i(\epsilon_0(k) + kv_0)^{1/2}/m
\end{array}
\right),\nonumber\\
\epsilon_0(k)=\sqrt{k^2v_0^2 + m^2},
\end{eqnarray}
one can finally obtain the diagonalized Hamiltonian,
\begin{eqnarray}
\label{eff_diagonal}
\hat{\cal H}_{H}^{\rm eff} &=&\sum_{|k|<\Lambda}
\Big[\epsilon_-(k)\tilde\eta_-^\dag(k)\tilde\eta_-(k)
+\epsilon_0(k)\tilde\eta_0^\dag(k)\tilde\eta_0(k)
\nonumber\\
&&
-\epsilon_+(k)\tilde\eta_+^\dag(k)\tilde\eta_+(k)\Big], %\nonumber\\
\end{eqnarray}
where $\epsilon_\mp(k)=\epsilon_0(k)\pm H$.
While three bands $\epsilon_{\pm,0}(k)$ are degenerate at $H=0$, a 
finite field $H$ mixes the first and third fermions $\xi_\nu^1$ and
$\xi_\nu^3$, and leaves these bands split. The dispersions 
$\epsilon_{\pm,0}(k)$ indicate that the Majorana
fermion theory correctly reproduces the Zeeman splitting. Fields 
$\tilde\eta_+$, $\tilde\eta_0^\dag$ and $\tilde\eta_-^\dag$ may 
respectively be regarded as $S^z=1,0$ and $=-1$ magnon creation operators.   
Here, let us consider how the symmetry operations are represented 
in the picture of the magnon fields $\tilde\eta_{\pm,0}$. 
Using the transformations (\ref{U(1)_Dirac_boson}) and 
(\ref{one-site_Dirac_boson}) and the relationship between 
($\psi_\nu$,$\xi_\nu^3$) and $\tilde\eta_{\pm,0}$ 
[Eqs.~(\ref{Fourier_fermion}) and (\ref{Bogoliu})],
one finds that the U(1) rotation along the spin z axis and the one-site
translation respectively correspond to   
\begin{eqnarray} 
\tilde\eta_\pm \to  e^{i\gamma}\tilde\eta_\pm, 
\label{eta_symm_U(1)}\\ 
\tilde\eta_\pm(k) \to -e^{ik a_0}  \tilde\eta_\pm(k), 
\hspace{0.5cm}
\tilde\eta_0(k)\to
\left\{
\begin{array}{ccc}
-e^{ik a_0} \tilde\eta_0(k)  &  (k>0)\\
-e^{-ik a_0} \tilde\eta_0(k) &  (k<0)
\end{array}
\right..
\label{eta_symm_trans}
\end{eqnarray}
%By using $\tilde\xi_{L,R}^3$, the final tranaformation in
%Eq.~(\ref{eta_symm_trans}) is simply rewritten as 
%$\tilde\xi_{L,R}^3(k)\to -e^{ik a_0}\tilde\xi_{L,R}^3(k)$. 
If we neglect the $k$-dependence of ${\cal U}_k$ or
consider only fermion fields around $k=0$, we could interpret that 
the site-parity transformation in Eq.~(\ref{parity_Dirac_boson}) 
is realized by 
\begin{eqnarray} 
\label{eta_symm_parity}
\tilde\eta_+(k)\to \mp i\tilde\eta_+(-k), \hspace{0.5cm} 
\tilde\eta_-(k)\to \pm i\tilde\eta_-(-k),\nonumber\\
\tilde\eta_0(k)\to
\left\{
\begin{array}{ccc}
\pm i\tilde\eta_0(-k) & (k>0)\\
\mp i\tilde\eta_0(-k) & (k<0)
\end{array}
\right..
\end{eqnarray}
The Hamiltonian (\ref{eff_diagonal}) is clearly invariant under these
transformations. One has to keep in mind that the symmetry
operation~(\ref{eta_symm_parity}) is valid only around $k=0$.

\begin{figure}[t]
\centerline{\scalebox{0.3}{\includegraphics{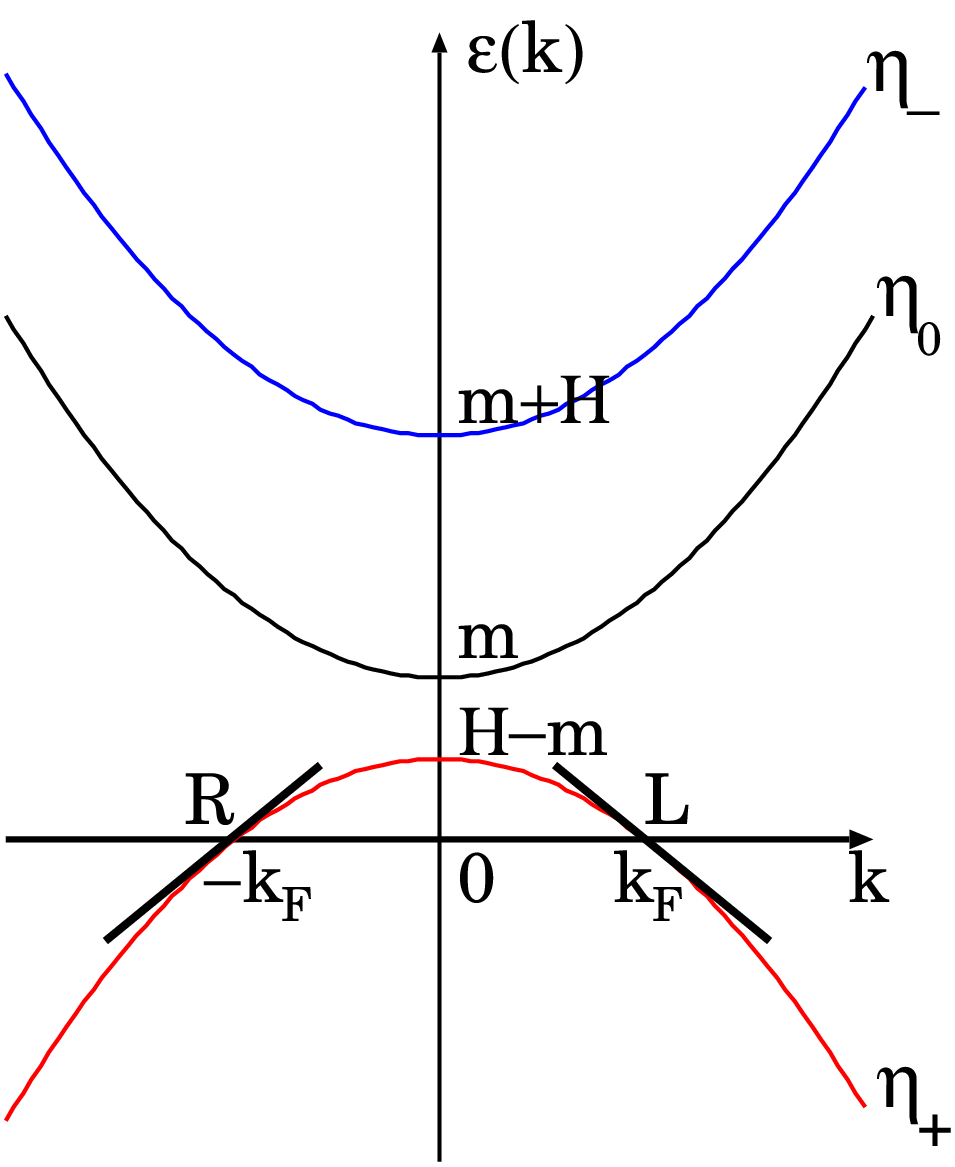}}}
\caption{\label{fig1} Spin-1 magnon band structure at $H>m$.}
\end{figure}

As $H$ exceeds the Haldane gap $m$, the band $\epsilon_+(k)$ crosses the
zero energy line. This just corresponds to the C-IC transition of
the spin-1 AF chain~(\ref{spin1AF})~\cite{Aff-bose,Aff-magBEC,Kon,Fath}. 
Then the system enters in a
critical TLL phase with the magnon $\tilde\eta_+$ condensed. The
two remaining bands $\epsilon_{-,0}(k)$ are still massive and completely
empty of magnons: $\langle\mu^3\rangle\neq 0$ and 
$\langle\sigma^3\rangle= 0$ still hold. Focusing on the
low-energy physics in this case of $H>m$, we can approximate the $S^z=1$
magnon field $\tilde\eta_+$ as a new Dirac fermion with a linear
dispersion as in Fig.~\ref{fig1}. The left and right movers 
$L(x)$ and $R(x)$ of the Dirac fermion are defined as  
\begin{eqnarray}
\label{new_Dirac}
L(x)=\frac{1}{\sqrt{L}}\sum_{|k|<\Lambda'}e^{ikx}\tilde L(k),&&\,\,\,\,
R(x)=\frac{1}{\sqrt{L}}\sum_{|k|<\Lambda'}e^{ikx}\tilde R(k),\nonumber\\
\tilde L(k)=\tilde\eta_+(k+k_F),&&\,\,\,\, \tilde R(k)=\tilde\eta_+(k-k_F),
\end{eqnarray}
where the Fermi wave number $k_F=\sqrt{H^2-m^2}/v_0$ is determined from 
$\epsilon_+(k_F)=0$, and the cut off $\Lambda'$ should be much smaller
than $\Lambda$. Using the Dirac fermion, one can rewrite the effective
Hamiltonian~(\ref{eff_diagonal}) in the case of $H>m$, 
in the real space, as follows: 
\begin{eqnarray}
\label{eff_TLL}
\hat{\cal H}_{H>m}^{\rm eff} &=& \int dx\,\Big[
iv_F\left(L^\dag\partial_x L - R^\dag\partial_x R \right)\nonumber\\
&&+\frac{i}{2}v_0\left(\xi_L^3\partial_x\xi_L^3-\xi_R^3\partial_x\xi_R^3\right)
+mi\xi_L^3\xi_R^3\nonumber\\
&&+\eta_-^\dag
\left(-\frac{v_0^2}{2m}\partial_x^2+m+H+{\cal O}(\partial_x^4)\right)
\eta_-\Big]\nonumber\\
&=& \hat{\cal H}[L,R]+\hat{\cal H}[\xi^3]+\hat{\cal H}[\eta_-],
\end{eqnarray}
where $v_F=\partial\epsilon_+/\partial k|_{k=k_F}=\sqrt{H^2-m^2}v_0/H$ is
the Fermi velocity, and
$\eta_-(x)=\frac{1}{\sqrt{L}}\sum_ke^{ikx}\tilde\eta_-(k)$.
Take notice that the linearized dispersion and the definition of $L$ and
$R$ become less reliable as the field $H$ becomes closer to the lower
critical value $m$ ($v_F\to 0$).

So far we have omitted the inter-magnon interaction terms in this
section. They would yield interactions among $L$, $R$, $\xi_\nu^3$ and
$\eta_-$ (see the next section). A better method of dealing with such
terms is the Abelian bosonization. Through it, the Dirac-fermion part in
the effective theory~(\ref{eff_TLL}) is mapped to the following 
Gaussian model,
\begin{eqnarray}
\label{Dirac_boso}
\hat{\cal H}[L,R]&\to&\hat{\cal H}[\phi]=\int dx\,\frac{v}{2}
\left[K(\partial_x\theta)^2+\frac{1}{K}(\partial_x\phi)^2\right],
\end{eqnarray}
where $\phi$ is a scalar field, $\theta$ is the dual of $\phi$, $v$ is
the renormalized Fermi velocity, and $K$ is the TLL parameter: 
although $K=1$ and $v=v_F$ at the stage of Eq.~(\ref{eff_TLL}), 
both parameters are subject to the renormalization 
due to irrelevant interaction effects neglected here.  
The Hamiltonian~(\ref{Dirac_boso}) just describes the low-energy physics
of the TLL phase in the chain~(\ref{spin1AF}).

%%%%%%%%%%%%%%%%%%%%%%%%%%%%%%%%%%%%%%%%%%%%%%%%%%%%%%%%
%%%%%%%%%%%%%%%%%%%%%%%%%%%%%%%%%%%%%%%%%%%%%%%%%%%%%%%%
%%%%%%%%%%%%%%%%%%%%%%%%%%%%%%%%%%%%%%%%%%%%%%%%%%%%%%%%
%%%%%%%%%%%%%%%%%%%%%%%%%%%%%%%%%%%%%%%%%%%%%%%%%%%%%%%%
%%%%%%%%%%%%%%%%%%%%%%%%%%%%%%%%%%%%%%%%%%%%%%%%%%%%%%%%
%%%%%%%%%%%%%%%%%%%%%%%%%%%%%%%%%%%%%%%%%%%%%%%%%%%%%%%%
%%%%%%%%%%%%%%%%%%%%%%%%%%%%%%%%%%%%%%%%%%%%%%%%%%%%%%%%
%%%%%%%%%%%%%%%%%%%%%%%%%%%%%%%%%%%%%%%%%%%%%%%%%%%%%%%%
%%%%%%%%%%%%%%%%%%%%%%%%%%%%%%%%%%%%%%%%%%%%%%%%%%%%%%%%
%%%%%%%%%%%%%%%%%%%%%%%%%%%%%%%%%%%%%%%%%%%%%%%%%%%%%%%%
%%%%%%%%%%%%%%%%%%%%%%%%%%%%%%%%%%%%%%%%%%%%%%%%%%%%%%%%
%%%%%%%%%%%%%%%%%%%%%%%%%%%%%%%%%%%%%%%%%%%%%%%%%%%%%%%%
%%%%%%%%%%%%%%%%%%%%%%%%%%%%%%%%%%%%%%%%%%%%%%%%%%%%%%%%
%%%%%%%%%%%%%%%%%%%%%%%%%%%%%%%%%%%%%%%%%%%%%%%%%%%%%%%%
%%%%%%%%%%%%%%%%%%%%%%%%%%%%%%%%%%%%%%%%%%%%%%%%%%%%%%%%

\section{Field-theory representation of spin operators in $\bol H>\bol m$}
\label{spin_op}
Making use of the contents in the preceding sections, we construct a 
field-theory representation of spin operators in the
uniform-field-induced TLL phase in the spin-1 AF chain~(\ref{spin1AF}).

\subsection{Spin uniform component and Magnetization}
\label{uniform_spin}
First, we consider the uniform component of the spin, 
$\vec J= \vec J_L+\vec J_R$ in Eq.~(\ref{spin_fermion}).
In the low-energy limit, Eqs.~(\ref{Bogoliu}) and
(\ref{new_Dirac}) allow us to approximate the ``old'' Dirac fermion
$\psi_\nu$ as follows:
\begin{eqnarray}
\label{Dirac_old_new}
\psi_R(x)&\approx& \frac{1}{\sqrt{2}}\eta_-(x)+e^{ik_Fx}U_+ L(x)
+e^{-ik_Fx}U_- R(x),\nonumber\\
\psi_L(x)&\approx& \frac{i}{\sqrt{2}}\eta_-(x)-ie^{ik_Fx}U_- L(x)
-ie^{-ik_Fx}U_+ R(x),
\end{eqnarray}
where the prefactor $1[i]/\sqrt{2}$ of the first term is the $(1,1)$
$[(2,1)]$ component of the matrix ${\cal U}_{k=0}$, 
and $U_\pm=m/\sqrt{2H(H\pm\sqrt{H^2-m^2})}$ is the $(1,1)$ component
of ${\cal U}_{\mp k_F}$. Note that
$U_\pm$ have properties 
(\rnum{1}) $U_\pm\to1/\sqrt{2}$ at $H\to m+0$, (\rnum{2}) 
$U_+^2+U_-^2=1$, and (\rnum{3}) $U_+U_-=\frac{m}{2H}$. 
From this relationship between $\psi_\nu$ and $(L,R,\eta_-)$ 
[see also Eqs.~(\ref{eta_symm_U(1)})-(\ref{eta_symm_parity})], we know
how the latter three fields are transformed by symmetry operations. 
The U(1) rotation in Eq.~(\ref{U(1)_Dirac_boson}) corresponds to 
\begin{eqnarray}
\label{U(1)_LReta}
\eta_-\to e^{i\gamma}\eta_-,\hspace{0.5cm}
\left\{
\begin{array}{cc}
L\to e^{i\gamma}L\\
R\to e^{i\gamma}R
\end{array}
\right..
\end{eqnarray}
The one-site translation in Eq.~(\ref{one-site_Dirac_boson}) is mapped to 
\begin{eqnarray}
\label{onesite_LReta}
\eta_-(x)\to -\eta_-(x+a_0),\hspace{0.5cm}
\left\{
\begin{array}{cc}
L(x)\to -e^{ik_Fa_0}L(x+a_0)\\
R(x)\to -e^{-ik_Fa_0}R(x+a_0)
\end{array}
\right..
\end{eqnarray}
The site-parity transformation in Eq.~(\ref{parity_Dirac_boson}), 
$\psi_L(x)\to\mp\psi_R(-x)$ and $\psi_R(x)\to\pm\psi_L(-x)$, can be
reproduced by  
\begin{eqnarray}
\label{parity_LReta}
\eta_-(x)\to \pm i\eta_-(-x),\hspace{0.5cm}
\left\{
\begin{array}{cc}
L(x)\to \mp i R(-x)\\
R(x)\to \mp i L(-x)
\end{array}
\right..
\end{eqnarray}
One can confirm that the Hamiltonian~(\ref{eff_TLL}) is invariant under
these transformations.  

%Unfortunately, we cannot find any simple mappings for
%$(L,R,\eta_-)$ that correspond to the site-parity transformation 
%in Eq.~(\ref{parity_Dirac_boson}), i.e., $\psi_{L,R}\to\psi_{R,L}$.
%This is because fields $(L,R,\eta_-)$ consist of 
%a complicated mixture of $\psi_L$ and $\psi_R$. (The site-parity
%symmetry will again be considered in Sec.~\ref{sym_inTLL}.) 

Using the relation~(\ref{Dirac_old_new}), we can 
rewrite fields $\xi_\nu^{1,3}$ in $\vec J_{L,R}$ in terms of
$L$, $R$, and $\eta_-$. As a result, the currents $\vec J_{L,R}$ are
reexpressed as 
\numparts
\begin{eqnarray}
\label{uniform_fermiform}
J_L^1(x) &\approx& \Big[-\frac{i}{2}(\eta_-^\dag+\eta_-)
+\frac{i}{\sqrt{2}}U_-(e^{-ik_Fx}L^\dag+e^{ik_Fx}L)\nonumber\\
&&+\frac{i}{\sqrt{2}}U_+(e^{ik_Fx}R^\dag + 
e^{-ik_Fx}R)\Big]\xi_L^3,
\label{uniform_L1}\\ %nonumber\\
J_L^2(x) &\approx& \Big[\frac{1}{2}(\eta_-^\dag-\eta_-)
-\frac{1}{\sqrt{2}}U_-(e^{-ik_Fx}L^\dag - e^{ik_Fx}L)\nonumber\\
&&-\frac{1}{\sqrt{2}}U_+(e^{ik_Fx}R^\dag -
 e^{-ik_Fx}R)\Big]\xi_L^3,
\label{uniform_L2}\\ %nonumber\\
J_L^3(x) &\approx& \frac{1}{4}(\eta_-\eta_-^\dag-\eta_-^\dag\eta_-)
+\frac{U_-^2}{2}(LL^\dag-L^\dag L)+\frac{U_+^2}{2}(RR^\dag-R^\dag R) 
\nonumber\\
&&+\frac{U_-}{\sqrt{2}}\left[e^{ik_Fx}\eta_-^\dag L + {\rm h.c}\right]
%e^{-ik_Fx}L^\dag\eta_-]
+\frac{U_+}{\sqrt{2}}\left[e^{-ik_Fx}\eta_-^\dag R +{\rm h.c}\right] 
%e^{ik_Fx}R^\dag\eta_-]
\nonumber\\
&&-U_+U_-\left[e^{-2ik_Fx}L^\dag R + {\rm h.c}\right],
%e^{2ik_Fx}R^\dag L],
\label{uniform_L3}\\ %nonumber\\
J_R^1(x) &\approx& \Big[\frac{1}{2}(\eta_-^\dag-\eta_-)
+\frac{1}{\sqrt{2}}U_+(e^{-ik_Fx}L^\dag - e^{ik_Fx}L)\nonumber\\
&&+\frac{1}{\sqrt{2}}U_-(e^{ik_Fx}R^\dag - 
e^{-ik_Fx}R)\Big]\xi_R^3,
\label{uniform_R1}\\ %nonumber\\
J_R^2(x) &\approx& \Big[\frac{i}{2}(\eta_-^\dag+\eta_-)
+\frac{i}{\sqrt{2}}U_+(e^{-ik_Fx}L^\dag + e^{ik_Fx}L)\nonumber\\
&&+\frac{i}{\sqrt{2}}U_-(e^{ik_Fx}R^\dag + 
e^{-ik_Fx}R)\Big]\xi_R^3,
\label{uniform_R2}\\ %nonumber\\
J_R^3(x) &\approx& \frac{1}{4}(\eta_-\eta_-^\dag-\eta_-^\dag\eta_-)
+\frac{U_+^2}{2}(LL^\dag-L^\dag L)+\frac{U_-^2}{2}(RR^\dag-R^\dag R) 
\nonumber\\
&&-\frac{U_+}{\sqrt{2}}\left[e^{ik_Fx}\eta_-^\dag L + {\rm h.c}\right]
%e^{-ik_Fx}L^\dag\eta_-]
-\frac{U_-}{\sqrt{2}}\left[e^{-ik_Fx}\eta_-^\dag R +{\rm h.c}\right] 
%e^{ik_Fx}R^\dag\eta_-]
\nonumber\\
&&-U_+U_-\left[e^{-2ik_Fx}L^\dag R + {\rm h.c}\right],
%e^{2ik_Fx}R^\dag L],
\label{uniform_R3} %nonumber\\
\end{eqnarray}
\endnumparts
It is easily found that the above $\vec J_{L,R}$ are 
appropriately transformed for the symmetry operations~(\ref{U(1)_LReta}), 
(\ref{U(1)_Dirac_boson}), (\ref{onesite_LReta}), 
(\ref{one-site_Dirac_boson}), (\ref{parity_LReta}) and (\ref{parity_Dirac_boson}).
In addition, the Hermitian nature of all current operators
is preserved in Eq.~(47). %(\ref{uniform_fermiform}).   
Even if, instead of Eq.~(\ref{Dirac_old_new}), 
we use precise representations of continuous fields in the Fourier space
such as Eqs.~(\ref{Fourier_fermion}) and (\ref{new_Dirac}), 
we can finally arrive at the same expression as
Eq.~(47). %~(\ref{uniform_fermiform}). 

The Fermi wave number $k_F$ is related with 
the uniform magnetization $M=\langle S_j^z\rangle$. It is
given by the formula~(47) or %~(\ref{uniform_fermiform}) or 
the Fourier-space representation of $J^3=J_L^3+J_R^3$. The result is   
\begin{eqnarray}
\label{Mag_kF}
M &=& \frac{1}{N}\left\langle \sum_j S_j^z\right\rangle = 
\frac{1}{N}\int dx\,\langle J^3(x)\rangle\nonumber\\
&&\left(\approx \frac{1}{N}\int dx 
\,\frac{1}{2}\left[
\langle L L^\dag-L^\dag L\rangle
+\langle R R^\dag-R^\dag R\rangle
+\langle \eta_-\eta_-^\dag\rangle
\right]\right)  \nonumber\\
&=&\sum_{|k|<k_F}\langle\tilde\eta_+(k)\tilde\eta_+^\dag(k)\rangle
=\frac{k_F a_0}{\pi}.
\end{eqnarray}
From the derivation process of Eq.~(\ref{Mag_kF}), 
we also find that the first three terms in Eqs.~(\ref{uniform_L3}) and 
(\ref{uniform_R3}) should respectively be regarded as  
\begin{eqnarray}
\label{L3_R3_replace}
\frac{1}{4}\left(\eta_-\eta_-^\dag-\eta_-^\dag\eta_-\right)
+\frac{U_-^2}{2}\left(LL^\dag-L^\dag L\right)
+\frac{U_+^2}{2}\left(RR^\dag-R^\dag R\right) 
\nonumber\\
\,\,\,\,\to \,\,-U_-^2:L^\dag L:-U_+^2:R^\dag R: +\frac{M}{2a_0}
-\frac{1}{2}\eta_-^\dag\eta_-,
\nonumber\\
\frac{1}{4}\left(\eta_-\eta_-^\dag-\eta_-^\dag\eta_-\right)
+\frac{U_+^2}{2}\left(LL^\dag-L^\dag L\right)
+\frac{U_-^2}{2}\left(RR^\dag-R^\dag R\right) \nonumber\\
\,\,\,\,\to\,\,-U_+^2:L^\dag L:-U_-^2:R^\dag R: +\frac{M}{2a_0}
-\frac{1}{2}\eta_-^\dag\eta_-.
\end{eqnarray}

Utilizing the results~(47) %~(\ref{uniform_fermiform})
-(\ref{L3_R3_replace}), 
and applying the Abelian bosonization to the Dirac fermion $(L,R)$, 
we can straightforwardly lead to the following partially 
bosonized currents $\vec J_{L,R}$: 
\begin{eqnarray}
\label{boson_currents}
J_L^1(x) &\approx & \Big[-\frac{i}{2}(\eta_-^\dag+\eta_-)
+\frac{iU_-}{\sqrt{\pi\alpha'}}\kappa_L
\cos(\sqrt{4\pi}\phi_L-\pi Mx/a_0)\nonumber\\
&&+\frac{iU_+}{\sqrt{\pi\alpha'}}\kappa_R
\cos(\sqrt{4\pi}\phi_R-\pi Mx/a_0)\Big]\xi_L^3,
\nonumber\\%%%%%%%%%
J_L^2(x) &\approx& \Big[\frac{1}{2}(\eta_-^\dag-\eta_-)
-\frac{iU_-}{\sqrt{\pi\alpha'}}\kappa_L
\sin(\sqrt{4\pi}\phi_L-\pi Mx/a_0)\nonumber\\
&&+\frac{iU_+}{\sqrt{\pi\alpha'}}\kappa_R
\sin(\sqrt{4\pi}\phi_R-\pi Mx/a_0)\Big]\xi_L^3,
\nonumber\\%%%%%%%%%
J_L^3(x) &\approx& 
-\frac{U_-^2}{\sqrt{\pi}}\,\partial_x\phi_L
-\frac{U_+^2}{\sqrt{\pi}}\,\partial_x\phi_R + \frac{M}{2a_0}
-\frac{1}{2}\eta_-^\dag\eta_-
\nonumber\\
&&+\frac{U_-}{\sqrt{4\pi\alpha'}}\left[\eta_-^\dag\kappa_L 
e^{-i\sqrt{4\pi}\phi_L+i\pi Mx/a_0} + {\rm h.c}\right]
\nonumber\\
&&+\frac{U_+}{\sqrt{4\pi\alpha'}}\left[\eta_-^\dag\kappa_R
e^{i\sqrt{4\pi}\phi_R-i\pi Mx/a_0} +{\rm h.c}\right] 
\nonumber\\
&&-i\frac{U_+U_-}{\pi\alpha'}\kappa_L\kappa_R
\sin(\sqrt{4\pi}\phi-2\pi Mx/a_0),
\nonumber\\%%%%%%%%%
J_R^1(x) &\approx & \Big[\frac{1}{2}(\eta_-^\dag-\eta_-)
+\frac{iU_+}{\sqrt{\pi\alpha'}}\kappa_L
\sin(\sqrt{4\pi}\phi_L-\pi Mx/a_0)\nonumber\\
&&-\frac{iU_-}{\sqrt{\pi\alpha'}}\kappa_R
\sin(\sqrt{4\pi}\phi_R-\pi Mx/a_0)\Big]\xi_R^3,
\nonumber\\%%%%%%%%%
J_R^2(x) &\approx & \Big[\frac{i}{2}(\eta_-^\dag+\eta_-)
+\frac{iU_+}{\sqrt{\pi\alpha'}}\kappa_L
\cos(\sqrt{4\pi}\phi_L-\pi Mx/a_0)\nonumber\\
&&+\frac{iU_-}{\sqrt{\pi\alpha'}}\kappa_R
\cos(\sqrt{4\pi}\phi_R-\pi Mx/a_0)\Big]\xi_R^3,
\nonumber\\%%%%%%%%%
J_R^3(x) &\approx& 
-\frac{U_+^2}{\sqrt{\pi}}\,\partial_x\phi_L
-\frac{U_-^2}{\sqrt{\pi}}\,\partial_x\phi_R + \frac{M}{2a_0}
-\frac{1}{2}\eta_-^\dag\eta_-
\nonumber\\
&&-\frac{U_+}{\sqrt{4\pi\alpha'}}\left[\eta_-^\dag\kappa_L 
e^{-i\sqrt{4\pi}\phi_L+i\pi Mx/a_0} + {\rm h.c}\right]
\nonumber\\
&&-\frac{U_-}{\sqrt{4\pi\alpha'}}\left[\eta_-^\dag\kappa_R
e^{i\sqrt{4\pi}\phi_R-i\pi Mx/a_0} +{\rm h.c}\right] 
\nonumber\\
&&-i\frac{U_+U_-}{\pi\alpha'}\kappa_L\kappa_R
\sin(\sqrt{4\pi}\phi-2\pi Mx/a_0),
\end{eqnarray}
where $\alpha'\sim1/\Lambda'$ is the short-distance cut off, and
$\kappa_{L(R)}$ is the Klein factor for the field $L(R)$. 
Therefore, the bosonized uniform components of the spin are written as 
%%%%%%%%%%%%%%%%%%  our first formula  %%%%%%%%%%%%%%%%%%%%%
\numparts
\begin{eqnarray}
\label{boson_uniform}
J^3(x)&\approx & \frac{M}{a_0}
-\frac{1}{\sqrt{\pi}}\partial_x\phi 
-\eta_-^\dag\eta_-
%\nonumber\\
%&&
-iC_0^z\kappa_L\kappa_R\sin(\sqrt{4\pi}\phi-2\pi Mx/a_0)
\nonumber\\
&&+C_1^z\left(\eta_-^\dag\kappa_L 
e^{-i\sqrt{\pi}(\phi+\theta)+i\pi Mx/a_0}+{\rm h.c}\right)
\nonumber\\
&&-C_1^z\left(\eta_-^\dag\kappa_R
e^{-i\sqrt{\pi}(\phi-\theta)-i\pi Mx/a_0}+{\rm h.c}\right)
+\cdots,\label{boson_uniform_1}
\\%%%%%%%%%%%%%%%%%%%%%%%%%%%%%%%%%
J^+(x)&\equiv&  J^1(x)+i J^2(x)
\nonumber\\
&\approx& \bigg\{-i\eta_- + e^{-i\sqrt{\pi}\theta}
\Big[iC_1 \kappa_L e^{-i\sqrt{\pi}\phi+\pi Mx/a_0}
\nonumber\\
&&+ i C_2 \kappa_R e^{i\sqrt{\pi}\phi-\pi Mx/a_0}
+\cdots\Big]+\cdots\bigg\}\xi_L^3
\nonumber\\
&&+\bigg\{-\eta_- - e^{-i\sqrt{\pi}\theta}
\Big[C_2\kappa_L e^{-i\sqrt{\pi}\phi+\pi Mx/a_0}
\nonumber\\
&&+C_1\kappa_R e^{i\sqrt{\pi}\phi-\pi Mx/a_0}+\cdots\Big]
+\cdots\bigg\}\xi_R^3,\label{boson_uniform_2}
\end{eqnarray}
\endnumparts
where $C_n$ and $C_n^z$ are nonuniversal constants, which contain the
effects of irrelevant terms neglected at the
stage~(\ref{eff_TLL}). When the irrelevant terms are not taken into
account, constants $C_0^z$, $C_1^z$, $C_1$, and 
$C_2$ are respectively proportional to $U_+U_-$, $U_- -U_+$, $U_-$,
and $U_+$. Therefore, it is inferred that $C_1^z\to 0$ at $H\to m+0$. 
In Eqs.~(\ref{boson_currents}) and (51), %(\ref{boson_uniform})
we used the formula~(\ref{boso_1}). If, instead of it, we use another 
formula~(\ref{boso_1_v2}), we can add more irrelevant terms to 
Eqs.~(\ref{boson_currents}) and (51). %(\ref{boson_uniform}). 
Namely, we may perform the replacement,
\begin{eqnarray}
\label{replace_boson}
e^{\pm i\sqrt{\pi}(\phi+\theta) \mp i\pi Mx/a_0} &\to& 
\sum_{n=0}^\infty \tilde{C}_n e^{\pm i(2n+1)(\sqrt{\pi}\phi-\pi Mx/a_0)}
e^{\pm i\sqrt{\pi}\theta},\nonumber\\
e^{\pm i\sqrt{\pi}(\phi-\theta) \mp i\pi Mx/a_0} &\to& 
\sum_{n=0}^\infty \tilde{C}_n e^{\pm i(2n+1)(\sqrt{\pi}\phi-\pi Mx/a_0)}
e^{\mp i\sqrt{\pi}\theta},
\end{eqnarray}
where $\tilde{C}_n$ is a nonuniversal
constant. Equation~(51) %~(\ref{boson_uniform}) 
is a main result in this paper. 
We expect that replacing the Klein factors in Eq.~(51) %~(\ref{boson_uniform}) 
with a constant is admitted in most situations. (A typical situation is 
when one investigates asymptotic features of spin correlation
functions. See Sec.~\ref{Asympto}.)

Finally, using the
results~(47)%(\ref{uniform_fermiform})
-(\ref{boson_currents}), 
let us investigate how the current-current interaction $\lambda$
term modifies the parameter $K$ of the effective
theory~(\ref{Dirac_boso}). Since the SU(2) spin-rotation
symmetry is reduced to the U(1) one in the case with a finite $H$, 
the term $-\lambda\vec J_L\vec J_R$ might be modified to 
$-\lambda_\perp(J_L^1 J_R^1+ J_L^2 J_R^2)-\lambda_\parallel J_L^3J_R^3$,
in which we assume that both $\lambda_\perp$ and $\lambda_\parallel$ are
positive. Equations~(47) %(\ref{uniform_fermiform}) 
and (\ref{L3_R3_replace}) 
lead to the following representation of the current-current
interaction~\cite{note_MS}:
\numparts
\begin{eqnarray}
\label{current_int_1}
J_L^3(x_+)J_R^3(x) &\approx& \left(U_-^2:R^\dag R:+U_+^2:L^\dag L:\right)
\left(U_+^2:R^\dag R:+U_-^2:L^\dag L:\right)
\nonumber\\
&&+U_+^2U_-^2\left(L^\dag R(x_+)R^\dag L(x)+R^\dag L(x_+)L^\dag R(x)\right)
\nonumber\\
&&-\frac{M}{2a_0}\left(:L^\dag L:+:R^\dag R:\right)
\nonumber\\
&&+\frac{U_+^2+U_-^2}{2}\eta_-^\dag\eta_-
\left(:L^\dag L:+:R^\dag R:\right)
\nonumber\\
&&-\frac{U_+U_-}{2}\left[\eta_-^\dag\eta_-
(L(x_+)L^\dag(x)+R(x_+)R^\dag(x))+{\rm h.c.}\right]
\nonumber\\
&&+\frac{U_-^2-U_+^2}{2}\left(\eta_-^\dag(x_+)\eta_-^\dag(x)LR
+{\rm h.c.}\right)
\nonumber\\
&&+\,\,\,{\rm oscillating\,\, terms}\,\,\,+\,\,\,{\rm others}, 
\end{eqnarray}
\begin{eqnarray}
\label{current_int_2}
J_L^1(x_+)J_R^1(x)+J_L^2(x_+)J_R^2(x) \approx
\nonumber\\
\hspace{3cm} iU_+U_-\left(L^\dag L-LL^\dag + R^\dag R-RR^\dag\right)\xi_L^3\xi_R^3
\nonumber\\
\hspace{3cm} +\,\,\,{\rm oscillating\,\, terms}\,\,\,+\,\,\,{\rm others}, 
\end{eqnarray}
\endnumparts
where we used $x_+=x+\delta$ ($\delta$ is a small parameter of 
${\cal O}(a_0)$), 
``oscillating terms'' have a factor $e^{\pm i n\pi Mx/a_0}$ ($n$:
integer), which are irrelevant except for the case that $M$ is close 
to a special commensurate value, and ``others'' are constructed by only
massive fields $(\xi_\nu^3,\eta_-)$. The Hamiltonian~(\ref{eff_TLL}) 
plus the above current-current interaction is regarded as a more accurate 
low-energy effective theory for the uniform-field-driven 
TLL phase. Integrating out the part of massive fields
$(\xi_\nu^3,\eta_-)$ in the effective theory~\cite{note_CumExp}, 
we obtain a Hamiltonian for the interacting Dirac fermion
$(L,R)$ that just corresponds to the Gaussian theory~(\ref{Dirac_boso}).  
In the Gaussian theory framework, the first and second terms of 
Eq.~(\ref{current_int_1}) directly contribute to the renormalization of 
the TLL parameter $K$ and the velocity $v$ as follows. 
The bosonized forms of these two terms are written as 
\numparts
\begin{eqnarray}
\label{K_v_RG}
\left(U_-^2:R^\dag R:+U_+^2:L^\dag L:\right)
\left(U_+^2:R^\dag R:+U_-^2:L^\dag L:\right)
\nonumber\\
\hspace{0.7cm}
\to\frac{1}{4\pi}(\partial_x\phi)^2
-\frac{(U_-^2-U_+^2)^2}{4\pi}(\partial_x\theta)^2
\nonumber\\
\hspace{1.cm}
+\frac{U_-^4-U_+^4}{4\pi}\left[\partial_x\phi\partial_x\theta 
-\partial_x\theta \partial_x\phi\right],\label{K_v_RG_1}
\\
U_+^2U_-^2\left(L^\dag R(x_+)R^\dag L(x)+R^\dag L(x_+)L^\dag R(x)\right)
\nonumber\\
\hspace{0.7cm}
\to\frac{U_+^2U_-^2}{(2\pi\alpha')^2}\left[e^{i\sqrt{4\pi}(\phi(x_+)-\phi(x))}
+e^{-i\sqrt{4\pi}(\phi(x_+)-\phi(x))}
\right]
\nonumber\\
\hspace{0.7cm}
\approx -\frac{m^2}{4\pi H^2}\left(\partial_x\phi\right)^2+{\rm
 const}+\cdots,
\label{K_v_RG_2}
\end{eqnarray}
\endnumparts
where we used $U_+^2+U_-^2=1$ and $U_+U_-=\frac{m}{2H}$, 
and assumed $\delta=\alpha'$. 
The final term in Eq.~(\ref{K_v_RG_1}) could be negligible or be replaced
with a constant: such a procedure is supported by the equal-time 
commutation relation $[\phi(x),\theta(y)]=-\frac{i}{2}{\rm sgn}(x-y)$. 
Consequently, two terms in Eq.~(54) %(\ref{K_v_RG_1}) 
lead to an additional boson kinetic term, 
\begin{eqnarray}
\label{boson_additional}
-\frac{1}{4\pi}(\lambda_\parallel-\frac{m^2}{H^2}\lambda_\perp)(\partial_x\phi)^2
+\frac{\lambda_\parallel}{4\pi}(U_-^2-U_+^2)^2(\partial_x\theta)^2,
\end{eqnarray}
for the $K=1$ Gaussian theory~(\ref{Dirac_boso})~\cite{note_theta}. 
This term is inclined to make the parameter $K$ 
increase, i.e., $K>1$, provided that $\lambda_\parallel$ 
is the same order as $\lambda_\perp$. 
The relation $K>1$ is consistent with the predictions of the numerical
calculation in Ref.~\cite{Fath} and the NLSM approach~\cite{Kon}.
(It is shown in Ref.~\cite{Fath} that $K$ is between 1.0 and 1.5 
in the field-induced TLL state.) 
Besides Eq.~(54), %(\ref{K_v_RG_1}), 
other terms in Eq.~(53) %~(\ref{current_int_1}) 
also yields a correction of $K$ and $v$, 
but it is expected that the property $K>1$ is maintained. 
From Eq.~(53), %~(\ref{current_int_1}),
one also finds that the current-current interaction does not generate
any relevant or marginal vertex operators such as
$\cos(\sqrt{4\pi}\phi)$. 
It means that the massless TLL phase survives even when the $\lambda$ 
interaction is taken into account. The third normal-ordered term in
Eq.~(\ref{current_int_1}) provides only a small correction of the
magnetization $M$, and can be absorbed into the Gaussian theory via
$\phi(x)\to\phi(x)-{\rm const}\times x$.

%%%%%%%%%%%%%%%%%%%%%%%%%%%%%%%%%%%%%%%%%%%%%%%%%%%%%%%%%%%%%%%%
%%%%%%%%%%%%%%%%%%%%%%%%%%%%%%%%%%%%%%%%%%%%%%%%%%%%%%%%%%%%%%%%
%%%%%%%%%%%%%%%%%%%%%%%%%%%%%%%%%%%%%%%%%%%%%%%%%%%%%%%%%%%%%%%%
%%%%%%%%%%%%%%%%%%%%%%%%%%%%%%%%%%%%%%%%%%%%%%%%%%%%%%%%%%%%%%%%
%%%%%%%%%%%%%%%%%%%%%%%%%%%%%%%%%%%%%%%%%%%%%%%%%%%%%%%%%%%%%%%%
%%%%%%%%%%%%%%%%%%%%%%%%%%%%%%%%%%%%%%%%%%%%%%%%%%%%%%%%%%%%%%%%
%%%%%%%%%%%%%%%%%%%%%%%%%%%%%%%%%%%%%%%%%%%%%%%%%%%%%%%%%%%%%%%%
%%%%%%%%%%%%%%%%%%%%%%%%%%%%%%%%%%%%%%%%%%%%%%%%%%%%%%%%%%%%%%%%
%%%%%%%%%%%%%%%%%%%%%%%%%%%%%%%%%%%%%%%%%%%%%%%%%%%%%%%%%%%%%%%%
%%%%%%%%%%%%%%%%%%%%%%%%%%%%%%%%%%%%%%%%%%%%%%%%%%%%%%%%%%%%%%%%

\subsection{Spin staggered component}
\label{staggered_spin}
Compared with the formula~(51) %(\ref{boson_uniform}) 
for the uniform component of the spin, 
the way of leading to a field-theory formula for the
staggered component $\vec N(x)$ is not straightforward,
because we have to express the vertex operators $e^{\pm i\sqrt{\pi}\Theta}$ 
and $e^{\pm i\sqrt{\pi}\Phi}$ in Eq.~(\ref{spin-stagg}) 
in terms of another boson $(\phi,\theta)$ language and the magnon $\eta_-$ one.
To this end, we apply the discussion in Ref.~\cite{Fu-Zh}.

First, using some results in the previous sections, let us naively 
speculate the relation between the ``old'' boson $(\Phi,\Theta)$ and the
``new'' one $(\phi,\theta)$.
The expression of $J^3(x)$ in Eqs.~(\ref{spin-uniform}) and 
(\ref{boson_uniform}) indicates the following relation 
between $\Phi$ and $\phi$, 
\begin{eqnarray}
\label{J3}
-\frac{1}{\sqrt{\pi}}\partial_x\Phi(x) &\approx& 
-\frac{1}{\sqrt{\pi}}\partial_x\phi(x) + M/a_0
\nonumber\\
&& -iC_0^z\kappa_L\kappa_R\,\sin(\sqrt{4\pi}\phi-2\pi Mx/a_0)+\cdots.  
\end{eqnarray}
Performing the integration with respect to $x$ in Eq.~(\ref{J3}), 
we obtain 
\begin{eqnarray}
\label{Phi-phi}
\Phi(x)&\approx& \phi(x)-\sqrt{\pi}Mx/a_0+\cdots,
\end{eqnarray}
where we dropped the sin term because it oscillates and then vanishes 
in the integration. From this, we can expect that
\begin{eqnarray}
\label{rough_old_new_1}
e^{\pm i\sqrt{\pi}\Phi} &\sim &e^{\pm i(\sqrt{\pi}\phi-\pi Mx/a_0)},
\end{eqnarray}
holds within a rough approximation. We next consider dual fields $\Theta$ and
$\theta$. 
The U(1) spin rotation in Eq.~(\ref{U(1)_LReta}) could be realized by 
\begin{eqnarray}
\label{newboson_U(1)}
\theta(x)&\to&\theta(x)-\frac{\gamma}{\sqrt{\pi}},
\end{eqnarray}
in the boson language. The comparison between this and 
Eq.~(\ref{U(1)_Dirac_boson}) implies the relation,   
\begin{eqnarray}
\label{rough_old_new_2}
e^{\pm i\sqrt{\pi}\Theta} &\sim &e^{\pm i\sqrt{\pi}\theta}.
\end{eqnarray}

In order to raise the validity of speculations~(\ref{rough_old_new_1}) and
(\ref{rough_old_new_2}), and obtain
a more appropriate relationship between $e^{\pm i\sqrt{\pi}\Phi(\Theta)}$
and $(\phi,\theta,\eta_-)$, a standard bosonization formula, 
\begin{eqnarray}
\label{old_boson_fermion}
\psi_L=\frac{\zeta_L}{\sqrt{2\pi\alpha}}e^{-i\sqrt{\pi}(\Phi+\Theta)},
\hspace{0.5cm}
\psi_R=\frac{\zeta_R}{\sqrt{2\pi\alpha}}e^{i\sqrt{\pi}(\Phi-\Theta)},
\end{eqnarray}
is available. From this formula, 
we rewrite vertex operators 
$e^{- i\sqrt{\pi}\Phi}$ and $e^{- i\sqrt{\pi}\Theta}$ as 
\numparts
\begin{eqnarray}
\label{vertex_fermion_1}
e^{- i\sqrt{\pi}\Phi}&=& \sqrt{2\pi\alpha}
\left[{\cal D} e^{i\sqrt{\pi}\Theta}\zeta_L\psi_L  
+ (1-{\cal D})  e^{-i\sqrt{\pi}\Theta}\zeta_R\psi_R^\dag \right],
\\
\label{vertex_fermion_2}
e^{- i\sqrt{\pi}\Theta}&=&\sqrt{2\pi\alpha}
\left[{\cal D} e^{i\sqrt{\pi}\Phi}\zeta_L\psi_L 
+ (1-{\cal D})  e^{-i\sqrt{\pi}\Phi}\zeta_R\psi_R \right],
\end{eqnarray}
\endnumparts
where $\cal D$ is an arbitrary constant~\cite{note_product}. 
Then, employing approximated results~(\ref{rough_old_new_1}),
(\ref{rough_old_new_2}), (\ref{Dirac_old_new}), and the bosonization formula 
$L,R\sim e^{\mp i\sqrt{4\pi}\phi_{L,R}}$ 
in the right-hand sides in Eq.~(62), %~(\ref{vertex_fermion_1}), 
we can arrive at a desirable representation of $e^{\pm i\sqrt{\pi}\Phi(\Theta)}$.
%Hereafter, we set ${\cal D} =1/2$.
This method was proposed in Ref.~\cite{Fu-Zh}.

Following the above idea, we obtain 
\numparts
\begin{eqnarray}
\label{vertex_new_old}
e^{- i\sqrt{\pi}\Theta} &\approx& \sqrt{\frac{\alpha}{4\alpha'}} 
e^{- i\sqrt{\pi}\theta} \bigg[
U_-\left(-i\zeta_L\kappa_L+\zeta_R\kappa_R\right) 
\nonumber\\
&& +U_+\left(-i\zeta_L\kappa_R e^{i\sqrt{4\pi}\phi- 2\pi Mx/a_0}
+\zeta_R\kappa_L e^{-i\sqrt{4\pi}\phi+ 2\pi Mx/a_0} \right)\bigg]
\nonumber\\ 
&&+\frac{\sqrt{\pi\alpha}}{2}\left(i\zeta_L e^{i\sqrt{\pi}\phi- \pi Mx/a_0}
+\zeta_R e^{-i\sqrt{\pi}\phi+ \pi Mx/a_0} \right)\eta_- 
\nonumber\\
&&  + \cdots,\label{vertex_new_old_theta}
\\%%%%%%%%%%%%%%%%%%%%%%%%%%%%%%%%%%%%%%
e^{- i\sqrt{\pi}\Phi} &\approx& \sqrt{\frac{\alpha}{4\alpha'}} \bigg[
U_-\left(-i\zeta_L\kappa_L +\zeta_R\kappa_R \right)
e^{ -i\sqrt{\pi}\phi + \pi Mx/a_0} 
\nonumber\\
&& +U_+\left(-i\zeta_L\kappa_R +\zeta_R\kappa_L \right)
e^{i\sqrt{\pi}\phi - \pi Mx/a_0} \bigg]
\nonumber\\
&&+\frac{\sqrt{\pi\alpha}}{2}\left(i\zeta_L\eta_- e^{i\sqrt{\pi}\theta}
+\zeta_R\eta_-^\dag e^{-i\sqrt{\pi}\theta}\right)
+\cdots,\label{vertex_new_old_phi}
\end{eqnarray}
\endnumparts
where we set ${\cal D} =1/2$, and $\kappa_{L(R)}$ is the Klein factor 
of the field $L(R)$~\cite{note_Commute}. It is found that the naive 
expectations~(\ref{rough_old_new_1}) and (\ref{rough_old_new_2}) are
qualitatively consistent with the result~(63). % ~(\ref{vertex_new_old}). 
From this formula, our target, the staggered component of the spin may
be represented as  
\numparts
%%%%%%%%%%%%%%   our second formula     %%%%%%%%%%%%%%%%%%%%%%%%%%%%%%
\begin{eqnarray}
\label{spin_staggered_boson}
N^3(x) &= &\sigma^3\cos(\sqrt{\pi}\Phi)
\nonumber\\ 
&\approx& \sigma^3\bigg[
D_1^z \cos\left(\sqrt{\pi}\phi-\pi Mx/a_0\right) 
\nonumber\\
&& + {D_1^z}' \left\{(i\zeta_L+\zeta_R)\eta_- e^{i\sqrt{\pi}\theta}+{\rm h.c.}\right\}
\bigg]%\nonumber\\
%&& 
+\cdots,\label{spin_staggered_boson_z}
\\%%%%%%%%%%%%%%%%%%%%%%%%%%%%%%%%%%%%%%%%%%%%%%
N^+(x) &\equiv& N^1(x)+i N^2(x) = i\mu^3 e^{-i\sqrt{\pi}\Theta}
\nonumber\\
&\approx&  \mu^3 e^{-i\sqrt{\pi}\theta}
\bigg[ D_1 +D_2\cos\left(\sqrt{4\pi}\phi - 2\pi Mx/a_0\right)+\cdots\bigg]
\nonumber\\
&&+D'_1 \mu^3\left(i\zeta_L e^{i\sqrt{\pi}\phi- \pi Mx/a_0}
+\zeta_R e^{-i\sqrt{\pi}\phi+ \pi Mx/a_0} \right)\eta_- 
%\sin\left(\sqrt{\pi}\phi - \pi Mx/a_0\right)\eta_-
\nonumber\\
&&+\cdots,\label{spin_staggered_boson_xy}
\end{eqnarray}
\endnumparts
where $D_n^{(z,\prime)}$ are nonuniversal
constants, which include effects of irrelevant terms, and we naively
omitted some of Klein factors $\zeta_\nu$ and $\kappa_\nu$. 
%(therefore, details of
%vertex-operator terms are not uniquely determined). 
We used the formula $L,R\sim e^{\mp i\sqrt{4\pi}\phi_{L,R}}$ in
Eqs.~(63) and (64). %~(\ref{vertex_new_old}) and (\ref{spin_staggered_boson}). 
Instead of that, applying the formula~(\ref{boso_1_v2}), 
we can introduce more irrelevant terms in Eqs.~(63) and (64) 
%~(\ref{vertex_new_old}) and (\ref{spin_staggered_boson}) 
[see the replacement~(\ref{replace_boson})].
For instance, we might replace $\cos(\sqrt{\pi}\phi-\pi Mx/a_0)$ in
Eq.~(\ref{spin_staggered_boson}) with 
\begin{eqnarray}
\label{replace_cos}
\sum_{n=0}^\infty
\tilde{D}_n\cos\left[(2n+1)(\sqrt{\pi}\phi-\pi Mx/a_0)\right],
\end{eqnarray}
where $\tilde{D}_n$ is a nonuniversal constant.

%%%%%%%%%%%%%%%%%%%%%%%%%%%%%%%%%%%%%%%%%%%%%%%%%%%%%%%%%%%%%%%%
%%%%%%%%%%%%%%%%%%%%%%%%%%%%%%%%%%%%%%%%%%%%%%%%%%%%%%%%%%%%%%%%
%%%%%%%%%%%%%%%%%%%%%%%%%%%%%%%%%%%%%%%%%%%%%%%%%%%%%%%%%%%%%%%%
%%%%%%%%%%%%%%%%%%%%%%%%%%%%%%%%%%%%%%%%%%%%%%%%%%%%%%%%%%%%%%%%
%%%%%%%%%%%%%%%%%%%%%%%%%%%%%%%%%%%%%%%%%%%%%%%%%%%%%%%%%%%%%%%%
%%%%%%%%%%%%%%%%%%%%%%%%%%%%%%%%%%%%%%%%%%%%%%%%%%%%%%%%%%%%%%%%%%
%%%%%%%%%%%%%%%%%%%%%%%%%%%%%%%%%%%%%%%%%%%%%%%%%%%%%%%%%%%%%%%%%%
%%%%%%%%%%%%%%%%%%%%%%%%%%%%%%%%%%%%%%%%%%%%%%%%%%%%%%%%%%%%%%%%%%
%%%%%%%%%%%%%%%%%%%%%%%%%%%%%%%%%%%%%%%%%%%%%%%%%%%%%%%%%%%%%%%%%%
%%%%%%%%%%%%%%%%%%%%%%%%%%%%%%%%%%%%%%%%%%%%%%%%%%%%%%%%%%%%%%%%%%

\subsection{Symmetries}
\label{sym_inTLL}
Through several considerations, we obtained a field-theory formula for spin
operators, Eqs.~(51) and (64). 
%~(\ref{boson_uniform}) and (\ref{spin_staggered_boson}).
Using the formula, transformations~(\ref{U(1)_LReta})-(\ref{parity_LReta}) and 
the effective Hamiltonians~(\ref{eff_TLL}) and
(\ref{Dirac_boso}), we can consider how the symmetries of the spin-1 AF 
chain~(\ref{spin1AF}) are represented 
in the partially-bosonized effective theory framework.

(\rnum{1}) As we discussed already, a U(1) rotation around the spin z axis, 
$S_j^+\to e^{i\gamma}S_j^+$, could be realized by 
\begin{eqnarray}
\label{U(1)_final}
\eta_-\to e^{i\gamma}\eta_-,\hspace{1cm}
\theta\to\theta-\frac{\gamma}{\sqrt{\pi}}.
\end{eqnarray}

(\rnum{2}) The one-site translation $\vec S_j\to \vec S_{j+1}$ may
correspond to 
\begin{eqnarray}
\label{onesite_final}
\eta_-(x)\to-\eta_-(x+a_0),\hspace{0.5cm}
\xi_{L,R}^3(x)\to-\xi_{L,R}^3(x+a_0),\nonumber\\
\left\{
\begin{array}{ccc}
\theta(x)\to\theta(x+a_0)+\sqrt{\pi}\\
\phi(x)\to\phi(x+a_0)-\sqrt{\pi}M
\end{array}
\right.,
\hspace{0.5cm}
\left\{
\begin{array}{ccc}
\sigma^3(x)\to-\sigma^3(x+a_0)\\
\mu^3(x)\to \mu^3(x+a_0)
\end{array}
\right..
\end{eqnarray}
These transformations cooperatively change the staggered 
factor $(-1)^j$ in front of $\vec N(x)$ into $(-1)^{j+1}$.

(\rnum{3}) The site-parity transformation $\vec S_j\to\vec S_{-j}$
%in Eq.~(\ref{parity_Dirac_boson}) and (\ref{parity_LReta})
might be regarded as~\cite{note_Klein}  
\begin{eqnarray}
\label{parity_final}
\eta_-(x)\to\pm i\eta_-(-x),\hspace{0.5cm}
\left\{
\begin{array}{ccc}
\xi_L^3(x)\to\mp \xi_R^3(-x)\\
\xi_R^3(x)\to\pm \xi_L^3(-x)
\end{array}
\right.,
\nonumber\\
\left\{
\begin{array}{ccc}
\kappa_L\to\mp i\kappa_R\\
\kappa_R\to\mp i\kappa_L\\
\phi(x) \to -\phi(-x)\\
\theta(x) \to \theta(-x)
\end{array}
\right.,\hspace{0.2cm}
\left\{
\begin{array}{ccc}
\sigma^3(x)\to\sigma^3(-x)\\
\mu^3(x)\to \mu^3(-x)
\end{array}
\right.,
\hspace{0.2cm}
\left\{
\begin{array}{ccc}
\zeta_L\to\mp\zeta_R\\
\zeta_R\to \pm\zeta_L
\end{array}
\right..
\end{eqnarray}
These symmetry operations 
might tell us which additional terms are allowed to be
present in formulas~(51) and (64): %%%%%%%
for example, the staggered component $N^3(x)$ may include 
$\sigma^3(\{(i\zeta_L+\zeta_R)\partial_x\eta_-\}^n
e^{in\sqrt{\pi}\theta}+{\rm h.c.})$, 
$\sigma^3\eta_-^\dag\eta_-$, etc ($n$: an integer). 
These terms are properly transformed for all the symmetry operations.

If we integrate out massive fields in the partition function and
replace all massive-field parts in spin operators with their
expectation values, the effective Hamiltonian becomes the Gaussian
model~(\ref{Dirac_boso}) and the reduced spin operators containing only 
massless boson fields are given by
\begin{eqnarray}
\label{reduced_formulas}
J^3(x)&\to & \frac{M}{a_0}-\frac{1}{\sqrt{\pi}}\partial_x\phi 
+ C_0^z \cos\left(\sqrt{4\pi}\phi-2\pi Mx/a_0\right) +\cdots,
\nonumber\\
N^+(x) &\to& e^{-i\sqrt{\pi}\theta}\langle \mu^3\rangle 
\left[D_1+D_2\cos\left(\sqrt{4\pi}\phi-2\pi Mx/a_0\right) +
 \cdots\right],
\nonumber\\
(J^+(x))^2 &\to&e^{-2i\sqrt{\pi}\theta}\Big[{\bar C}_1+{\bar C}_2
\cos\left(\sqrt{4\pi}\phi-2\pi Mx/a_0\right)+\cdots\Big],
\nonumber\\
(N^3(x))^2 &\to&{\bar D}_1+{\bar D}_2
\cos\left(\sqrt{4\pi}\phi-2\pi Mx/a_0\right)+\cdots,
\end{eqnarray}
where ${\bar C}_n$ and ${\bar D}_n$ are nonuniversal constants, all the
Klein factors are naively eliminated, and 
we used $(\sigma^3(x))^2\sim {\rm const}$, 
$\langle\eta_-\eta_-^\dag\rangle\sim {\cal O}(a_0^{-1})$, and 
$\langle\xi_\nu^3\xi_{\nu'}^3\rangle\sim {\cal O}(a_0^{-1})$ 
(see ~\ref{FermiCorrlation}). This result indicates that the
compactification radius of $\phi$ and that of $\theta$ are respectively
$1/\sqrt{4\pi}$ and $1/\sqrt{\pi}$. In this bosonized-theory framework, 
we could interpret that the U(1) rotation $S_j^+\to e^{i\gamma}S_j^+$, 
the one-site translation $\vec S_j\to \vec S_{j+1}$, 
and the site-parity operation $\vec S_j\to\vec S_{-j}$,
respectively, corresponds to
\numparts 
\begin{eqnarray}
\label{reduced_symm}
\theta\to\theta-\frac{\gamma}{\sqrt{\pi}},
\label{reduced_symm_1}\\
\theta(x)\to\theta(x+a_0)+\sqrt{\pi},
\hspace{1cm}
\phi(x)\to\phi(x+a_0)-\sqrt{\pi}M,
\label{reduced_symm_2}\\
\phi(x) \to -\phi(-x),%+\sqrt{\pi}/2,
\hspace{1cm}
\theta(x) \to \theta(-x).\label{reduced_symm_3}
\end{eqnarray}
\endnumparts
The symmetries of Eq.~(70) %~(\ref{reduced_symm}) 
strongly restrict the emergence of the relevant vertex
operators in the effective field theory~(\ref{Dirac_boso}): 
the U(1) symmetry (\ref{reduced_symm_1}) 
[the translational symmetry (\ref{reduced_symm_2})] 
forbids $e^{i n\sqrt{\pi}\theta}$ [$e^{in\sqrt{4\pi}\phi}$] to exist 
in the effective Hamiltonian~\cite{OYA}. 
Consequently, the TLL state stably remains. 

One can see that the above transformations~(70) %~(\ref{reduced_symm}) 
are very similar to those of the effective bosonization theory 
for spin-1/2 AF chains (see~\ref{XXZ}).

%%%%%%%%%%%%%%%%%%%%%%%%%%%%%%%%%%%%%%%%%%%%%%%%%%%%%%%%
%%%%%%%%%%%%%%%%%%%%%%%%%%%%%%%%%%%%%%%%%%%%%%%%%%%%%%%%
%%%%%%%%%%%%%%%%%%%%%%%%%%%%%%%%%%%%%%%%%%%%%%%%%%%%%%%%
%%%%%%%%%%%%%%%%%%%%%%%%%%%%%%%%%%%%%%%%%%%%%%%%%%%%%%%%
%%%%%%%%%%%%%%%%%%%%%%%%%%%%%%%%%%%%%%%%%%%%%%%%%%%%%%%%
%%%%%%%%%%%%%%%%%%%%%%%%%%%%%%%%%%%%%%%%%%%%%%%%%%%%%%%%
%%%%%%%%%%%%%%%%%%%%%%%%%%%%%%%%%%%%%%%%%%%%%%%%%%%%%%%%
%%%%%%%%%%%%%%%%%%%%%%%%%%%%%%%%%%%%%%%%%%%%%%%%%%%%%%%%
\subsection{Asymptotic behavior of spin correlation functions}
\label{Asympto}
Applying the formulas~(51) and (64), %%%%%
let us investigate equal-time spin correlation functions 
in the field-induced TLL phase. 
The asymptotic forms of equal-time two-point functions of 
$\vec J$ and $\vec N$ are evaluated as 
\numparts
\begin{eqnarray}
\label{2point_JandN}
\langle J^3(x)J^3(0)\rangle &\approx& \left(\frac{M}{a_0}\right)^2
+ \frac{1}{\pi}\langle\partial_x\phi(x)\partial_x\phi(0)\rangle 
\nonumber\\
&&+{\cal A}^z \cos\left(2\pi M\frac{x}{a_0}\right)\langle e^{-i\sqrt{4\pi}\phi(x)}
e^{i\sqrt{4\pi}\phi(0)}\rangle +\cdots
\nonumber\\
&=& \left(\frac{M}{a_0}\right)^2 -\frac{K}{2\pi^2}\frac{1}{x^2}
\nonumber\\
&&+{\cal A}^z \cos\left(2\pi M\frac{x}{a_0}\right)
\left(\frac{\alpha'}{x}\right)^{2K} +\cdots,
\\%%%%%%%%%%%%%%%%%%%%%%%%%%%%%%%%%%%%%%%%
\langle J^+(x)J^-(0)\rangle &\approx& {\cal A}
\cos\left(\pi M\frac{x}{a_0}\right)
\langle e^{-i\sqrt{\pi}(\phi\pm\theta)(x)}
e^{i\sqrt{\pi}(\phi\pm\theta)(0)}\rangle 
\nonumber\\
&&\times \langle  \xi_{L/R}^3(x)\xi_{L/R}^3(0)\rangle +\cdots
\nonumber\\
&\sim & \cos\left(\pi M\frac{x}{a_0}\right)
\left(\frac{\alpha'}{x}\right)^{\frac{K+K^{-1}}{2}}
K_1\left(\frac{x}{\xi_c}\right)+\cdots,
\\%%%%%%%%%%%%%%%%%%%%%%%%%%%%%%%%%%%%%%%
\langle N^3(x)N^3(0)\rangle &\approx& {\cal B}^z
\cos \left(\pi M\frac{x}{a_0}\right)
\langle\sigma^3(x)\sigma^3(0)\rangle 
\nonumber\\
&&\times \langle e^{-i\sqrt{\pi}\phi(x)}e^{i\sqrt{\pi}\phi(0)}\rangle
+\cdots\nonumber\\
&\sim& \cos \left(\pi M\frac{x}{a_0}\right)
\left(\frac{\alpha'}{x}\right)^{K/2}
K_0\left(\frac{x}{\xi_c}\right)+\cdots,
\\%%%%%%%%%%%%%%%%%%%%%%%%%%%%%%%%%%%
\langle N^+(x)N^-(0)\rangle &\approx& {\cal B}_1
\langle\mu^3(x)\mu^3(0)\rangle 
\langle e^{-i\sqrt{\pi}\theta(x)}e^{i\sqrt{\pi}\theta(0)}\rangle\Big[1
\nonumber\\
&&+{\cal B}_2\cos\left(2\pi M\frac{x}{a_0}\right)
\langle e^{-i\sqrt{4\pi}\phi(x)}e^{i\sqrt{4\pi}\phi(0)}\rangle
+\cdots\Big]
\nonumber\\
&\sim& \left(\frac{\alpha'}{x}\right)^{1/(2K)}
+{\cal B}_2'\cos\left(2\pi M\frac{x}{a_0}\right)
\nonumber\\
&&\times\left(\frac{\alpha'}{x}\right)^{\frac{1}{2K}+2K}+\cdots,
\end{eqnarray}
\endnumparts
where ${\cal A}^{(z)}$ and ${\cal B}^{(z,\prime)}_{(n)}$ are 
nonuniversal constants, $K_\nu$ is the modified Bessel function, and
$x\gg \xi_c(=v_0/m)$. In the calculation of Eq.~(71), %%%%
we used Eqs.~(\ref{Ising_corr}) [see the comment~\cite{note2}], 
(\ref{corr-func}), 
(\ref{xi_2point})-(\ref{eta_2point}). From this result, 
longitudinal and transverse spin-spin correlators are determined as 
\numparts
\begin{eqnarray}
\label{asymp_spin_corr}
\langle S_j^zS_0^z\rangle &\approx&  M^2
-\frac{K}{2\pi^2}\left(\frac{a_0}{x}\right)^2 +{\cal C}_0^z 
\cos\left(2\pi Mj\right)\left(\frac{a_0}{x}\right)^{2K} 
\nonumber\\
&&+{\cal D}_1^z \cos \left(\pi (1+M)j\right)
\left(\frac{a_0}{x}\right)^{(K+1)/2}e^{-x/\xi_c}+\cdots,
\\%%%%%%%%%%%%%%%%%%%%%%%%%%%%%%%%%%%%%%%%%%%
\langle S_j^+ S_0^-\rangle &\approx& 
{\cal C}_1 \cos\left(\pi Mj\right)
\left(\frac{a_0}{x}\right)^{(K+K^{-1}+1)/2}e^{-x/\xi_c}
\nonumber\\
&&+{\cal D}_1  (-1)^j  %\cos\left(\pi\frac{x}{a_0}\right)
\left(\frac{a_0}{x}\right)^{1/(2K)}
\nonumber\\
&&+{\cal D}_2 \cos\left(\pi(1+2M)j\right)
\left(\frac{a_0}{x}\right)^{\frac{1}{2K}+2K}+\cdots,
\end{eqnarray}
\endnumparts
where $x=ja_0\gg \xi_c$, and ${\cal C}_q^{(z)}$ and 
${\cal D}_q^{(z)}$ are nonuniversal constants which are related to 
$C_n^{(z)}$ and $D_n^{(z,\prime)}$ in Eqs.~(51) and (64). %%%%
The first two terms in $\langle S_j^zS_0^z\rangle$ and 
the second term in $\langle S_j^+ S_0^-\rangle$ can also be 
derived by the NLSM plus Ginzbrug-Landau approach~\cite{Aff-magBEC,Kon},
but it is difficult to obtain all the other terms within the same
approach. One can verify that the
critical exponent of the incommensurate part around wave number 
$p\sim \pm 2k_F$ ($k_F=\pi M/a_0$) in $\langle S_j^zS_0^z\rangle$, 
$\eta_z=2K$, and that of the staggered part in 
$\langle S_j^+ S_0^-\rangle$, $\eta=1/(2K)$, satisfy the famous
relation $\eta\eta_z=1$~\cite{Hal2,S-T,Aff-magBEC,Fu-Zh,H-F}. 
In addition, it is found that the contribution around 
$p\sim \pi/a_0 \pm k_F$ in $\langle S_j^zS_0^z\rangle$ 
and that around $p\sim\pm k_F$ in $\langle S_j^+ S_0^-\rangle$ 
exhibit an exponential decay. 
Comparing Eq.~(72) with the spin
correlators of a two-leg spin-1/2 ladder in a uniform field in 
Ref.~\cite{Fu-Zh} would be instructive (although our calculations 
in Eqs.~(71) and (72) are rougher than those in Ref.~\cite{Fu-Zh}). 
Besides Eq.~(72), using the formulas~(51) and (64), one can calculate
various physical quantities (susceptibilities, dynamical structure
factors, NMR relaxation rates, etc) in the TLL phase~\cite{Kon}.

In all the calculations of this subsection, 
we assumed that the three systems 
$\hat{\cal H}[\phi]$, $\hat{\cal H}[\xi^3]$, and $\hat{\cal H}[\eta_-]$
are independent of each other. Although small interactions among these systems 
would actually be present, it is expected that their effects in 
the low-energy, long-distance physics are almost negligible and could be
absorbed into some parameters such as $K$, $v_{(0)}$, $m$, etc.

%%%%%%%%%%%%%%%%%%%%%%%%%%%%%%%%%%%%%%%%%%%%%%%%%%%%%%%%%%%%%%%%
%%%%%%%%%%%%%%%%%%%%%%%%%%%%%%%%%%%%%%%%%%%%%%%%%%%%%%%%%%%%%%%%
%%%%%%%%%%%%%%%%%%%%%%%%%%%%%%%%%%%%%%%%%%%%%%%%%%%%%%%%%%%%%%%%
%%%%%%%%%%%%%%%%%%%%%%%%%%%%%%%%%%%%%%%%%%%%%%%%%%%%%%%%%%%%%%%%
%%%%%%%%%%%%%%%%%%%%%%%%%%%%%%%%%%%%%%%%%%%%%%%%%%%%%%%%%%%%%%%%
%%%%%%%%%%%%%%%%%%%%%%%%%%%%%%%%%%%%%%%%%%%%%%%%%%%%%%%%%%%%%%%%%%
%%%%%%%%%%%%%%%%%%%%%%%%%%%%%%%%%%%%%%%%%%%%%%%%%%%%%%%%%%%%%%%%%%
%%%%%%%%%%%%%%%%%%%%%%%%%%%%%%%%%%%%%%%%%%%%%%%%%%%%%%%%%%%%%%%%%%
%%%%%%%%%%%%%%%%%%%%%%%%%%%%%%%%%%%%%%%%%%%%%%%%%%%%%%%%%%%%%%%%%%

\section{Some applications of the field-theory representation of spin 
operators}
\label{Application}
In this section, utilizing the derived
formulas~(51) and (64), %~(\ref{boson_uniform}) and (\ref{spin_staggered_boson}).
we discuss some topics for the low-energy physics in/around 
the uniform-field-induced TLL phase.

\subsection{String order parameter}
\label{string}
As a quantity characterizing the Haldane phase ($H<m$), there is the
nonlocal string order parameter which can detect the so-called 
``hidden AF long-range order'' in the phase~\cite{Nijs,Ke-Ta,Mag}. 
It is defined by 
\begin{eqnarray}
\label{string}
O_\alpha(i,j)&=&-\lim_{|i-j|\to\infty}\left\langle
S_i^\alpha \exp\left(i\pi\sum_{n=i+1}^{j-1}S_n^\alpha\right)S_j^\alpha
\right\rangle.
\end{eqnarray}
In the continuous-field-theory framework, we can predict~\cite{Naka,MS05_1} 
that the string parameter is approximated as follows: 
\begin{eqnarray}
\label{string_Ising}
O_\alpha(i,j)&\sim& \langle \mu^{\alpha+1}(x_i)\mu^{\alpha+2}(x_i)
\mu^{\alpha+1}(x_j)\mu^{\alpha+2}(x_j)\rangle,
\end{eqnarray}
where $x_i=ia_0$ and $x_j=ja_0$. Actually, the right-hand side is 
a finite, non-zero value in the Haldane phase 
where $\langle\mu^\alpha\rangle \neq 0$.

From the formula~(\ref{boso_3}), 
the z component of the string parameter is rewritten as 
\begin{eqnarray}
\label{z_string_Ising}
O_z(i,j)&\sim & 
\langle \cos(\sqrt{\pi}\Phi(x_i))\cos(\sqrt{\pi}\Phi(x_j))
\rangle.
\end{eqnarray}
In the field-induced TLL phase ($H>m$), 
the vertex operator $\exp(\pm i\sqrt{\pi}\Phi)$ is 
represented by using fields $(\phi,\theta,\eta_-)$ [see
Eqs.~(\ref{vertex_new_old_phi}) and (\ref{spin_staggered_boson})]. 
Hence, it is predicted that 
in the TLL phase, $O_z(i,j)$ behaves as 
\begin{eqnarray}
\label{z_string_Ising}
O_z(i,j)&\sim & 
\langle \cos(\sqrt{\pi}\phi(x_i)-\pi Mi)\cos(\sqrt{\pi}\phi(x_j)-\pi Mj)
\rangle+\cdots\nonumber\\
&\sim& \cos\left[\pi M(i-j)\right]
\left(\frac{a_0}{x_i-x_j}\right)^{K/2}+\cdots.
\end{eqnarray}
Namely, $O_z(i,j)$ is shown to exhibit power decay in the TLL phase.

%%%%%%%%%%%%%%%%%%%%%%%%%%%%%%%%%%%%%%%%%%%%%%%%%%
%%%%%%%%%%%%%%%%%%%%%%%%%%%%%%%%%%%%%%%%%%%%%%%%%%
%%%%%%%%%%%%%%%%%%%%%%%%%%%%%%%%%%%%%%%%%%%%%%%%%%
%%%%%%%%%%%%%%%%%%%%%%%%%%%%%%%%%%%%%%%%%%%%%%%%%%
%%%%%%%%%%%%%%%%%%%%%%%%%%%%%%%%%%%%%%%%%%%%%%%%%%
\subsection{SU(2)-invariant perturbations}
\label{SU(2)per}
In Secs.~\ref{SU(2)per}-\ref{U1sym_break}, 
we investigate typical perturbations for the critical TLL state. 
In particular, we focus on whether or not the perturbation terms 
yield a first excitation gap, and the symmetries of them.

In this subsection, we discuss two terms:  
the bond alternation $\sum_j(-1)^j J\delta \vec S_j\cdot\vec S_{j+1}$ 
($|\delta|\ll 1$) and the next-nearest-neighbor (NNN) exchange 
$\sum_j J_2 \vec S_j\cdot\vec S_{j+2}$ ($|J_2|\ll J$),
which are invariant under the global SU(2) transformation.  
Because a global U(1) symmetry, a part of the SU(2) one, is usually 
necessary for the realization of the TLL i.e., a 
$c=1$ CFT~\cite{Gogo,Tsv,CFT,Mag}, 
and (as we mentioned in Sec.~\ref{sym_inTLL}) it prohibits all vertex 
operators with the dual field $\theta(x)$ from 
emerging in the effective Hamiltonian, 
we expect, without any calculations, that the TLL phase 
survives even when these perturbations are applied. Let us consider the
two terms below in more detail.

From the continuous-field formula~(\ref{spin_fermion}), 
we can expect that the bond alternation term is approximated as  
\begin{eqnarray}
\label{bond_alt}
\sum_j(-1)^j J\delta \vec S_j\cdot\vec S_{j+1} = \sum_jJ\delta (-1)^j 
\left[\frac{1}{2}\left(S_j^+S_{j+1}^-+{\rm h.c}\right)+S_j^zS_{j+1}^z\right]
\nonumber\\
\sim J\delta a_0 \int dx
\bigg\{(-1)^j\Big[
\frac{1}{2}\left(J^+(x)J^-(x+a_0)+{\rm h.c}\right)+J^z(x)J^z(x+a_0)
\nonumber\\
-\frac{1}{2}\left(N^+(x)N^-(x+a_0)+{\rm h.c}\right)+N^z(x)N^z(x+a_0)
\Big]\nonumber\\
+\frac{1}{2}\left(-J^+(x)N^-(x+a_0)+N^+(x)J^-(x+a_0)+{\rm h.c}\right)
\nonumber\\
-J^z(x)N^z(x+a_0) +N^z(x)J^z(x+a_0)\bigg\}.
\end{eqnarray}
Let us substitute the formulas~(51) and (64) %%% 
into the above result, 
although such a procedure sometimes causes mistakes 
(see the comment~\cite{note_product}). As a result, we obtain 
\begin{eqnarray}
\label{alt_boso}
\sum_j(-1)^j J\delta \vec S_j\cdot\vec S_{j+1} &\to& J\delta \int dx
\sum_{n=1}^\infty \Big[\lambda_n'\cos(n(\sqrt{4\pi}\phi-2\pi Mj)+\alpha_n')
\nonumber\\
&&+(-1)^j\lambda_n\cos(n(\sqrt{4\pi}\phi-2\pi Mj)+\alpha_n)\Big]\nonumber\\
&&+\cdots,
\end{eqnarray}
where $\lambda_n^{(\prime)}$ and $\alpha_n^{(\prime)}$ are 
nonuniversal constants. Here, we used $(\sigma^3(x))^2\sim {\rm const}$,
and the operator product expansion (OPE) $\mu^3\times \xi_{L,R}^3\sim
\sigma^3+\cdots$~\cite{Gogo,Tsv,CFT,Fran}, 
and then integrated out all the massive-field parts 
in the partition function (see the comment~\cite{note_CumExp}). 
In order to more restrict the form of Eq.~(\ref{alt_boso}), we utilize 
the symmetry argument. For the one-site translation and 
the site-parity operation, the bond alternation term changes its
sign. Its bosonized form should also have
the same property; namely, we require the right-hand side 
in Eq.~(\ref{alt_boso}) to change the sign 
for $\phi(x)\to\phi(x+a_0)-\sqrt{\pi}M$ and 
$\phi(x)\to -\phi(-x)$ [see Eq.~(70)]. %\ref{reduced_symm}
Consequently, we set $\lambda_n'=0$ and $\alpha_n=\pi/2$. The
resultant form of Eq.~(\ref{alt_boso}) is 
\begin{eqnarray}
\label{bnd_alt_final}
\sum_j(-1)^j J\delta \vec S_j\cdot\vec S_{j+1} &\to& J\delta \int dx
\sum_{n=1}^\infty 
(-1)^j\lambda_n\sin(n(\sqrt{4\pi}\phi-2\pi Mj))
\nonumber\\
&&+\cdots. 
\end{eqnarray}
This bosonized form indicates that in general, 
the bond alternation is irrelevant in the TLL phase due to the
staggered factor $(-1)^j$ and the phase $2n\pi Mj$. 
However, when $M=1/2$ (half of the saturation), 
a relevant interaction $\sin(\sqrt{4\pi}\phi)$ originates 
from the $n=1$ term in Eq.~(\ref{bnd_alt_final}) 
because of the cancellation of two factors $(-1)^j$ and $2\pi Mj$. 
The scaling dimension $\Delta_s$ of $\sin(\sqrt{4\pi}\phi)$ is $K$ 
($1\leq K<1.5$ \cite{Fath}). At this case of $M=1/2$, 
the low-energy physics may be described by a sine-Gordon theory, 
and an infinitesimal bond alternation induces a
finite excitation gap and a finite dimerization parameter 
$\langle \vec S_{2n}\cdot\vec S_{2n+1}-
\vec S_{2n+1}\cdot\vec S_{2n+2}\rangle$. 
The existence of the gap further means that 
the bond alternation brings an $M=1/2$ plateau in the uniform
magnetization process. 
The scaling argument near a criticality~\cite{Cardy} shows that 
the bond-alternation-induced gap $\Delta_\delta$ and 
dimerization parameter respectively behave as 
\begin{eqnarray}
\label{bond_gap_dimer}
\Delta_\delta \sim |\delta|^{1/(2-K)}, \nonumber\\%\hspace{0.5cm}
\langle \vec S_{2n}\cdot\vec S_{2n+1}
-\vec S_{2n+1}\cdot\vec S_{2n+2}\rangle 
\sim  -{\rm sgn}(\delta) |\delta|^{K/(2-K)},
\end{eqnarray}
for a small $|\delta|$. %(The symbol Sgn denotes the sign function.) 
Because of the inequality $1/(2-K)>1$, 
the gap gradually grows with increasing $|\delta|$. 
These predictions for the (small) bond alternation are consistent with
previous numerical~\cite{Tone} and analytical~\cite{Totsuka} works.

Next, let us tern to the NNN coupling ($J_2$) term. Through an 
argument similar to that above, we arrive in the following result: 
\begin{eqnarray}
\label{NNN_boso}
\sum_j J_2 \vec S_j\cdot\vec S_{j+2}&\to& J_2\int dx
\Big\{\frac{a_0}{\pi}(\partial_x\phi)^2-\frac{2M}{\sqrt{\pi}}\partial_x\phi
\nonumber\\ 
&& +\sum_{n=1}^\infty \tilde\lambda_n 
\cos(n(\sqrt{4\pi}\phi-2\pi Mj))+\cdots\Big\},
\end{eqnarray}
where $\tilde\lambda_n$ is a nonuniversal constant. We used the symmetry
argument: the right-hand side in Eq.~(\ref{NNN_boso}) is invariant
under $\phi(x)\to\phi(x+a_0)-\sqrt{\pi}M$ and $\phi(x)\to -\phi(-x)$.
Since the bosonized NNN coupling does not contain any relevant operators
for arbitrary magnetization values, we conclude that the TLL phase
remains even when a sufficiently small NNN exchange perturbation 
is introduced. The derivative term $\partial_x\phi$ is absorbed into the
Gaussian part via $\phi\to\phi-{\rm const}J_2\times x$ and provides a small
correction of the magnetization $M$. 
While the boson quadratic term $(\partial_x\phi)^2$ makes
the velocity $v$ and the TLL parameter $K$ modify. After easy
calculations, we can see that when $J_2>0$ ($J_2<0$), $M$ and $K$
decrease (increase), but $v$ increases (decreases).

%%%%%%%%%%%%%%%%%%%%%%%%%%%%%%%%%%%%%%%%%%%%%%%%%%
%%%%%%%%%%%%%%%%%%%%%%%%%%%%%%%%%%%%%%%%%%%%%%%%%%
%%%%%%%%%%%%%%%%%%%%%%%%%%%%%%%%%%%%%%%%%%%%%%%%%%
%%%%%%%%%%%%%%%%%%%%%%%%%%%%%%%%%%%%%%%%%%%%%%%%%%
%%%%%%%%%%%%%%%%%%%%%%%%%%%%%%%%%%%%%%%%%%%%%%%%%%
\subsection{Axially symmetric terms}
\label{U1sym}
In this subsection, we consider three kinds of U(1)-symmetric
perturbation terms: the single-ion anisotropy $D_z\sum_j(S_j^z)^2$ 
(so-called $D$ term), the XXZ type anisotropy 
$J\Delta_z\sum_j S_j^z S_{j+1}^z$, and the staggered-field Zeeman term 
along the spin z axis $-h_z\sum_j(-1)^jS_j^z$
($|D_z|,|h_z|\ll J$ and $|\Delta_z|\ll 1$). 
As in the cases of the bond alternation and the NNN coupling, 
there is a high possibility that the TLL phase survives 
as these perturbations are added.

Following the similar argument to that in the last subsection, 
we can bosonize the three terms as follows: 
\numparts
\begin{eqnarray}
\label{U(1)perturbation_boso}
D_z\sum_j(S_j^z)^2 &\to& D_z\int dx 
\Big\{\frac{a_0}{\pi}(\partial_x\phi)^2-\frac{2M}{\sqrt{\pi}}\partial_x\phi
\nonumber\\ 
&& +\sum_{n=1}^\infty d_n 
\cos(n(\sqrt{4\pi}\phi-2\pi Mj))+\cdots\Big\},%\nonumber
\label{U(1)perturbation_boso_1}\\
J\Delta_z\sum_j S_j^z S_{j+1}^z&\to& J\Delta_z\int dx 
\Big\{\frac{a_0}{\pi}(\partial_x\phi)^2-\frac{2M}{\sqrt{\pi}}\partial_x\phi
\nonumber\\ 
&& +\sum_{n=1}^\infty \delta_n 
\cos(n(\sqrt{4\pi}\phi-2\pi Mj))+\cdots\Big\},%\nonumber
\label{U(1)perturbation_boso_2}\\
-h_z\sum_j(-1)^jS_j^z &\to & -h_z\int dx \Big\{(-1)^j\sum_{n=1}^\infty
z_n \cos(n(\sqrt{4\pi}\phi-2\pi Mj))\nonumber\\
&&+\cdots\Big\},\label{U(1)perturbation_boso_3}
\end{eqnarray}
\endnumparts
where $d_n$, $\delta_n$ and $z_n$ are nonuniversal constants. 
For instance, we required the bosonization form 
(\ref{U(1)perturbation_boso_3}) of the staggered-field term 
to change its sign for the one-site translation 
$\phi(x)\to\phi(x+a_0)-\sqrt{\pi}M$ and to be invariant under the
site-parity transformation $\phi(x)\to -\phi(-x)$. 
The results (\ref{U(1)perturbation_boso_1}) and 
(\ref{U(1)perturbation_boso_2}) suggest that the $D$ term and the XXZ
exchange play almost the same roles in the low-energy, long-distance
physics of the TLL phase.  
If $D_z>0$ ($D_z<0$), $M$ and $K$ decrease (increase), while $v$
increases (decreases). As expected, these three terms generally do not destroy
the TLL state. However, as in the case of the bond alternation term, 
when the uniform magnetization $M$ becomes close to 1/2, the
staggered-field term involves a relevant term $\cos(\sqrt{4\pi}\phi)$ 
with $\Delta_s=K$. Therefore, a staggered-field-induced gap 
$\Delta_{h_z}$ opens and $\langle S_j^z\rangle$ obtains a staggered
component at $M=1/2$. From the standard scaling argument, 
the gap and the magnetization are shown to behave as 
\begin{eqnarray}
\label{stag_gap_mag}
\Delta_{h_z} &\sim& |h_z|^{1/(2-K)}, \nonumber\\
\langle S_j^z\rangle &\approx& 
\frac{1}{2} + (-1)^j f\left(\frac{H}{J}\right) 
{\rm sgn}(h_z)|h_z|^{K/(2-K)},
\end{eqnarray}
where $f$ is a function.

%%%%%%%%%%%%%%%%%%%%%%%%%%%%%%%%%%%%%%%%%%%%%%%%%%
%%%%%%%%%%%%%%%%%%%%%%%%%%%%%%%%%%%%%%%%%%%%%%%%%%
%%%%%%%%%%%%%%%%%%%%%%%%%%%%%%%%%%%%%%%%%%%%%%%%%%
%%%%%%%%%%%%%%%%%%%%%%%%%%%%%%%%%%%%%%%%%%%%%%%%%%
%%%%%%%%%%%%%%%%%%%%%%%%%%%%%%%%%%%%%%%%%%%%%%%%%%
\subsection{Axial-symmetry-breaking terms}
\label{U1sym_break}
Here, we discuss three axial-symmetry-breaking terms: 
the $D$ term with the x component of spin $D_x\sum_j(S_j^x)^2$, 
another kind of the single-ion anisotropy (so-called $E$ term) 
$E\sum_j[(S_j^x)^2-(S_j^y)^2]=\frac{E}{2}\sum_j[(S_j^+)^2+(S_j^-)^2]$, 
and the staggered-field term along the spin x axis 
$-h_x\sum_j(-1)^j S_j^x$ ($|D_x|,|E|,|h_x|\ll J$). 
Since the three terms destroy the axial U(1) symmetry
($\theta\to\theta-\gamma/\sqrt{\pi}$), vertex
operators with $\theta(x)$ are allowed to be present in the effective
Hamiltonian. It is inferred that
such vertex operators cause an instability of the TLL state, and then 
a finite excitation gap occurs.
Here, note that the $D_x$ and $E$ terms are invariant under the $\pi$
rotation $S^+_j\to-S^+_j$ ($\theta\to
\theta-\sqrt{\pi}$)~\cite{Ess-Aff}, whereas the
$h_x$ term obtains a minus sign via the same rotation. 
Furthermore, the $\pi/2$ rotation leaves the $E$ term change the sign.

Through some calculations, the three perturbation terms are bosonized as 
\numparts
\begin{eqnarray}
\label{Dx_boso}
D_x\sum_j(S_j^x)^2 &\to& D_x\int dx
\Big\{\cos(\sqrt{4\pi}\theta)\sum_{n=0}^\infty d_n^{(1)}
\cos(n(\sqrt{4\pi}\phi-2\pi Mj))
\nonumber\\
&&+\cos(4\sqrt{\pi}\theta)\sum_{n=0}^\infty  d_n^{(2)}
\cos(n(\sqrt{4\pi}\phi-2\pi Mj))
\nonumber\\
&&+\sum_{n=0}^\infty d_n^{(3)}
\cos(n(\sqrt{4\pi}\phi-2\pi Mj))+\cdots\Big\},
\end{eqnarray}
\begin{eqnarray}
\label{E_boso}
E\sum_j[(S_j^x)^2-(S_j^y)^2] &\to & E \int dx
\Big\{\cos(\sqrt{4\pi}\theta)\times
\nonumber\\
&&\sum_{n=0}^\infty e_n
\cos(n(\sqrt{4\pi}\phi-2\pi Mj))+\cdots\Big\},
\end{eqnarray}
\begin{eqnarray}
\label{hx_boso}
-h_x\sum_j(-1)^j S_j^x &\to & -h_x \int dx
\Big\{\cos(\sqrt{\pi}\theta)
\nonumber\\
&&\times \sum_{n=0}^\infty 
x_n \cos(n(\sqrt{4\pi}\phi-2\pi Mj))
+\cdots\Big\},
\end{eqnarray}
\endnumparts
where $d_n^{(l)}$, $e_n$ and $x_n$ are nonuniversal constants. 
(Since the bosonized form of the anisotropic exchange 
$J\Delta_x\sum_j S_j^x S_{j+1}^x$ is the same type as 
Eq.~(\ref{Dx_boso}), we do not discuss it here.) 
One should note the following properties:  
(\rnum{1}) $\cos(\sqrt{\pi}\theta)\to-\cos(\sqrt{\pi}\theta)$, 
$\cos(\sqrt{4\pi}\theta)\to\cos(\sqrt{4\pi}\theta)$ 
and $\cos(4\sqrt{\pi}\theta)\to\cos(4\sqrt{\pi}\theta)$ for the $\pi$
rotation, and (\rnum{2}) 
$\cos(\sqrt{4\pi}\theta)\to-\cos(\sqrt{4\pi}\theta)$ for the $\pi/2$
rotation.

The most relevant operator in both $D_x$ and $E$ terms is always 
$\cos(\sqrt{4\pi}\theta)$ with $\Delta_s=1/K<2$.  
Thus, the low-energy properties can be explained by a
sine-Gordon model, and a gap emerges. Supposing $d_0^{(1)}$ and
$e_0$ are positive, the potential $\cos(\sqrt{4\pi}\theta)$ 
pins the phase field $\theta$ to 
$\pm\sqrt{\pi}/2$ ($0$ or $\sqrt{\pi}$) modulo $2\sqrt{\pi}$ 
for $D_x,E>0$ ($D_x,E<0$). In such a case of 
$\theta\to \pm\sqrt{\pi}/2$ ($\to 0$ or $\sqrt{\pi}$), 
it is expected that $\langle S^x_j \rangle \sim 
(-1)^j\langle\cos(\sqrt{\pi}\theta)\rangle=0$ ($\neq 0$) and 
$\langle S^y_j \rangle \sim 
(-1)^j\langle\sin(\sqrt{\pi}\theta)\rangle \neq 0$ ($=0$). From this
prediction and the scaling argument [$\Delta_s$ of
$e^{i\sqrt{\pi}\theta}$ is $1/(4K)$], we conclude that 
for a small $D_x$ term, the gap $\Delta_{D_x}$ and the transverse 
component of the spin moment increase as follows: 
\begin{eqnarray}
\label{D_x_scal}
\Delta_{D_x}\sim |D_x|^{K/(2K-1)},
\nonumber\\
\left\{
\begin{array}{ccc}
\langle S^x_j \rangle & \sim & 0 \\
\langle S^y_j \rangle & \sim & (-1)^j {D_x}^{1/(8K-4)}
\end{array}
\right.,
\hspace{0.5cm} ({\rm for}\,\,D_x>0),\nonumber\\
\left\{
\begin{array}{ccc}
\langle S^x_j \rangle & \sim & (-1)^j |D_x|^{1/(8K-4)}\\
\langle S^y_j \rangle & \sim &  0
\end{array}
\right.,
\hspace{0.5cm} ({\rm for}\,\,D_x<0).
\end{eqnarray}
Of course, for a small $E$ term, the similar results hold: we may
replace $D_x$ to $E$ in Eq.~(\ref{D_x_scal}). The staggered moment
(i.e., a N\'eel order) along the spin x or y axis shows that the $D_x$
(or $E$) term causes the spontaneous breakdown of the one-site
translational symmetry.  
Because of $K/(2K-1)<1$ and $1/(8K-4)<1$, the $D_x$- or $E$-term-induced
gap and the transverse moment $\langle S^{x,y}_j \rangle$ rapidly
increases with the growth of $|D_x|$ or $|E|$.

The staggered-field $h_x$ term contains the relevant term
$\cos(\sqrt{\pi}\theta)$ with $\Delta_s=1/(4K)<2$. Therefore, 
the $h_x$-induced gap $\Delta_{h_x}$ and the staggered magnetization 
are shown to behave as   
\begin{eqnarray}
\label{h_x_scal}
\Delta_{h_x}\sim |h_x|^{4K/(8K-1)},\hspace{0.5cm}
\langle S^x_j \rangle\sim (-1)^j{\rm sgn}(h_x)|h_x|^{1/(8K-1)}.
\end{eqnarray}
For the spin-1/2 AF Heisenberg chain, which low-energy sector is also
described by a TLL theory (see Introduction), 
a staggered field also yields a gap and a staggered moment. 
Oshikawa and Affleck~\cite{O-A} show that in the spin-1/2 case, 
$\Delta_{h_x}\sim |h_x|^{2/3}$ and 
$\langle S^x_j \rangle \sim (-1)^j{\rm sgn}(h_x)|h_x|^{1/3}$. 
The result~(\ref{h_x_scal}) thus indicates that 
the growth of both the gap and the staggered moment 
in the spin-1 case is much sharper than that in the spin-1/2 case.

%%%%%%%%%%%%%%%%%%%%%%%%%%%%%%%%%%%%%%%%%%%%%%%%%%%%%%%%
%%%%%%%%%%%%%%%%%%%%%%%%%%%%%%%%%%%%%%%%%%%%%%%%%%%%%%%%
%%%%%%%%%%%%%%%%%%%%%%%%%%%%%%%%%%%%%%%%%%%%%%%%%%%%%%%%
%%%%%%%%%%%%%%%%%%%%%%%%%%%%%%%%%%%%%%%%%%%%%%%%%%%%%%%%
%%%%%%%%%%%%%%%%%%%%%%%%%%%%%%%%%%%%%%%%%%%%%%%%%%%%%%%%
%%%%%%%%%%%%%%%%%%%%%%%%%%%%%%%%%%%%%%%%%%%%%%%%%%%%%%%%

\subsection{Magnon decay induced by axial-symmetry-breaking terms}
\label{decay}
In this subsection, we briefly mention the magnon-decay processes, 
which have already in some detail discussed in Ref.~\cite{Ess-Aff}.
In order to consider such processes, let us go back to the effective
Hamiltonian~(\ref{eff_diagonal}), where $\tilde\eta_+$,
$\tilde\eta_0^\dag$ and $\tilde\eta_-^\dag$ respectively 
denote the $S^z=+1$, $0$ and $-1$ magnon creation operators. 
The marginally irrelevant
$\lambda$ term, omitted in Eq.~(\ref{eff_diagonal}), 
just contribute to the magnon decay.

First, we focus on the U(1)-symmetric AF chain (\ref{spin1AF}) 
without any perturbations. 
Since the one-magnon excitations are present only around
$p=\pi/a_0$, only the decay from a magnon to an odd number of magnons 
is possible. If $H$ is increased so that $3\epsilon_+(0)<\epsilon_-(0)$ 
($H>m/2$) [$3\epsilon_+(0)<\epsilon_0(0)$ ($H>2m/3$)] are satisfied, 
a magnon $\tilde\eta_-^\dag$ [$\tilde\eta_0^\dag$] is energetically
permitted to decay into three magnons
$\tilde\eta_+\tilde\eta_+\tilde\eta_+$ via the $\lambda$ term. 
However, since $\tilde\eta_+$, $\tilde\eta_0^\dag$ and 
$\tilde\eta_-^\dag$ possess different eigenvalues of $S^z$ (namely,
three kinds of one-magnon states are in different sectors of the 
Hilbert space), this type of the decay is forbidden. Indeed, 
for a U(1) rotation $S_j^+\to e^{i\gamma}S_j^+$, 
$\tilde\eta_-$ [$\tilde\eta_0$] is transformed as 
$\tilde\eta_- \to e^{i\gamma}\tilde\eta_-$ 
[$\tilde\eta_0 \to \tilde\eta_0$ (invariant)], 
while the product $\tilde\eta_+\tilde\eta_+\tilde\eta_+$ obey a 
different rotation $\tilde\eta_+\tilde\eta_+\tilde\eta_+\to 
e^{3i\gamma}\tilde\eta_+\tilde\eta_+\tilde\eta_+$. Three
kinds of magnons ($\tilde\eta_+$, $\tilde\eta_0^\dag$,
$\tilde\eta_-^\dag$) therefore would be well-defined 
quasiparticles in the U(1)-symmetric system~(\ref{spin1AF}).

On the other hand, 
when a U(1)-symmetry-breaking term is introduced, 
there is a possibility that the above decay processes are allowed. 
In the case with the $D_x$ or $E$ terms, the 
$\pi$-rotation symmetry ($S_j^+\to -S_j^+$: $\gamma=\pi$), 
a part of the U(1) rotation, survives. For this rotation,  
$\tilde\eta_-$ and $\tilde\eta_+\tilde\eta_+\tilde\eta_+$ are odd,
whereas $\tilde\eta_0$ is even. Furthermore, from 
Eqs.~(\ref{eta_symm_trans}) and (\ref{eta_symm_parity}), we find 
that for example a sum of four terms 
``$\tilde\eta_-(k_1)\tilde\eta_+(k_2)\tilde\eta_+(-k_1)\tilde\eta_+(-k_2)
-(k_{1,2}\to-k_{1,2}) +{\rm h.c}$'' is invariant under both 
the one-site translation and the site-parity operation. (Because the
site-parity operation~(\ref{eta_symm_parity}) is the result of the
approximation, the requirement of the invariance under this
transformation might be too strong.) 
%$\tilde\eta_-^\dag(k)\to-e^{-ika_0}\tilde\eta_-^\dag(k)$ and 
%$\tilde\eta_+(k)\tilde\eta_+(-k)\tilde\eta_+(-k)\to
%-e^{-ika_0}\tilde\eta_+(k)\tilde\eta_+(-k)\tilde\eta_+(-k)$ for the
%one-site translation, and 
%$\tilde\eta_-^\dag(k)\to\mp i\tilde\eta_-^\dag(-k)$ and 
%$\tilde\eta_+(k)\tilde\eta_+(-k)\tilde\eta_+(-k)\to
%\mp i\tilde\eta_+(-k)\tilde\eta_+(k)\tilde\eta_+(k)$.
As a result, the process 
$\tilde\eta_-^\dag\to\tilde\eta_+\tilde\eta_+\tilde\eta_+$ is 
permissible. It is hence inferred that as sufficiently strong 
$D_x$ or $E$ terms are present in the system, 
$\tilde\eta_-^\dag$ magnons become ill-defined particles, 
and we should eliminate the magnon fields from the
Hamiltonian~(\ref{eff_TLL}) and the field-theory formulas 
of the spin, Eqs.~(51) and (64). 
In the case with the $h_x$ staggered-field term, symmetries of the $\pi$
rotation and the one-site translation are also broken (the two-site
translational symmetry remains).  
Therefore, the magnon decay would be more promoted. 
%there emerges a probability that another decay process 
%$\tilde\eta_-^\dag \to \tilde\eta_+\tilde\eta_+\tilde\eta_+$ occurs. 
When $H$ is further increased and the $\tilde\eta_+$-magnon 
condensation occurs ($H>m$), other types of the magnon decay are
energetically admitted: for instance, 
$\tilde\eta_-^\dag\to\tilde\eta_0^\dag\tilde\eta_0^\dag
\tilde\eta_+\tilde\eta_+\tilde\eta_+$.

From the simple discussion above, 
one sees that as $H$ is sufficiently strong, a large
axial-symmetry-breaking perturbation tends to make the 
lifetime of massive magnons shorten. However, if such
a perturbation is small enough, our effective theory framework would
still be reliable and have the ability to explain various low-energy
properties of spin-1 AF chains. 
%can predict various properties of spin-1 AF chains. 

%%%%%%%%%%%%%%%%%%%%%%%%%%%%%%%%%%%%%%%%%%%%%%%%%%%%%%%%
%%%%%%%%%%%%%%%%%%%%%%%%%%%%%%%%%%%%%%%%%%%%%%%%%%%%%%%%
%%%%%%%%%%%%%%%%%%%%%%%%%%%%%%%%%%%%%%%%%%%%%%%%%%%%%%%%
%%%%%%%%%%%%%%%%%%%%%%%%%%%%%%%%%%%%%%%%%%%%%%%%%%%%%%%%
%%%%%%%%%%%%%%%%%%%%%%%%%%%%%%%%%%%%%%%%%%%%%%%%%%%%%%%%
%%%%%%%%%%%%%%%%%%%%%%%%%%%%%%%%%%%%%%%%%%%%%%%%%%%%%%%%
%%%%%%%%%%%%%%%%%%%%%%%%%%%%%%%%%%%%%%%%%%%%%%%%%%%%%%%%
%%%%%%%%%%%%%%%%%%%%%%%%%%%%%%%%%%%%%%%%%%%%%%%%%%%%%%%%
%%%%%%%%%%%%%%%%%%%%%%%%%%%%%%%%%%%%%%%%%%%%%%%%%%%%%%%%
%%%%%%%%%%%%%%%%%%%%%%%%%%%%%%%%%%%%%%%%%%%%%%%%%%%%%%%%

\section{Summary}
\label{Summary}
In this paper, based on the Majorana fermion theory, 
we have reconsidered the field theory description of the spin-1 AF
chain~(\ref{spin1AF}), and derived an explicit field-theory 
form of spin operators in the uniform-field-driven TLL phase in the
chain~(\ref{spin1AF}), i.e., Eqs.~(51) and (64) [the corresponding
effective Hamiltonian is Eqs.~(\ref{eff_TLL}) and (\ref{Dirac_boso})]. 
%~(\ref{boson_uniform}) and (\ref{spin_staggered_boson}). 
From the formula, we have completely determined the
asymptotic forms of spin correlation functions (Sec.~\ref{Asympto}). 
Furthermore, applying the formula, we have investigated the string order
parameter and effects of some perturbation terms
(the bond alternation, the next-nearest interaction, anisotropy terms) in 
Sec.~\ref{Application}. We have estimated the excitation gaps and some
physical quantities (staggered moments and the dimerization parameter) 
generated from the perturbations. 
From Sec.~\ref{fermion_review} to Sec.~\ref{Application}, 
we have often argued how symmetries of the spin-1 AF
chain are represented in the effective field theory 
world.

Our results, especially Eqs.~(51) and (64), must be useful in
analyzing and understanding various spin-1 AF chains with magnons
condensed (i.e., with a finite magnetization) and 
the extended models of them (e.g., spin-1 AF ladders, spatially anisotropic
2D or 3D spin-1 AF systems). 
The results of Sec.~\ref{Application} guarantee this expectation. 
%demonstrate that the formula of spin operators is very useful in
%studying and understanding the low-energy physics 
%of/around the TLL phase of the spin-1 AF chain~(\ref{spin1AF}).
We will apply the contents of this paper to other spin-1 AF systems 
in the near future.

Determining nonuniversal coefficients 
in Eqs.~(51), (64) and (\ref{reduced_formulas}), 
especially those in front of terms including only
massless bosons, is important for more
quantitative predictions of (quasi) 1D spin-1 systems.
A powerful way of the determination is to accurately evaluate the 
long-distance behavior of spin correlation functions by means of a
numerical method such as DMRG and QMC~\cite{H-F}.

%Real spin-1 compounds such as NDMAP usually have anisotropy terms;
%single-ion anisotropy, dipole-dipole interaction, etc. 

%%%%%%%%%%%%%%%%%%%%%%%%%%%%%%%%%%%%%%%%%%%%%%%%%%%%%%%%
%%%%%%%%%%%%%%%%%%%%%%%%%%%%%%%%%%%%%%%%%%%%%%%%%%%%%%%%
%%%%%%%%%%%%%%%%%%%%%%%%%%%%%%%%%%%%%%%%%%%%%%%%%%%%%%%%
%%%%%%%%%%%%%%%%%%%%%%%%%%%%%%%%%%%%%%%%%%%%%%%%%%%%%%%%
%\section*{Aknowlefgements}
\ack
This work is supported by a Grant-in-Aid for Scientific 
Research (B) (No. 17340100) from the Ministry of Education, 
Culture, Sports Science and Technology of Japan.

%%%%%%%%%%%%%%%%%%%%%%%%%%%%%%%%%%%%%%%%%%%%%%%%%%%%%%%%
%%%%%%%%%%%%%%%%%%%%%%%%%%%%%%%%%%%%%%%%%%%%%%%%%%%%%%%%
%%%%%%%%%%%%%%%%%%%%%%%%%%%%%%%%%%%%%%%%%%%%%%%%%%%%%%%%
%%%%%%%%%%%%%%%%%%%%%%%%%%%%%%%%%%%%%%%%%%%%%%%%%%%%%%%%
%%%%%%%%%%%%%%%%%%%%%%%%%%%%%%%%%%%%%%%%%%%%%%%%%%%%%%%%
%%%%%%%%%%%%%%%%%%%%%%%%%%%%%%%%%%%%%%%%%%%%%%%%%%%%%%%%
%%%%%%%%%%%%%%%%%%%%%%%%%%%%%%%%%%%%%%%%%%%%%%%%%%%%%%%%
%%%%%%%%%%%%%%%%%%%%%%%%%%%%%%%%%%%%%%%%%%%%%%%%%%%%%%%%

\appendix
\section{Abelian bosonization for fermion systems}
\label{Abelian}
Here, we briefly summarize the Abelian bosonization.
As mentioned in Sec.~\ref{fermion_review}, in (1+1)D case, 
a massless Dirac (complex) fermion or two species of massless Majorana 
(real) fermions (critical Ising models) are 
equivalent to a massless bosonic Gaussian theory. 
The former Hamiltonian is written as 
\begin{eqnarray}
\label{Dirac_Majorana}
\hat{\cal H}[\psi] &=& \int \,\,dx\,\, 
ic\left(\psi_L^\dag\partial_x\psi_L-\psi_R^\dag\partial_x\psi_R\right)
\nonumber\\
&=&\int \,dx\,\,
\sum_{q=1,2}
\frac{i}{2}c\left(\xi_L^q\partial_x\xi_L^q-\xi_R^q\partial_x\xi_R^q\right)
=\hat{\cal H}[\xi^1,\xi^2] , 
\end{eqnarray}
where $\psi_\nu$ and $\xi_\nu^q$ are respectively the chiral
components of the Dirac fermion and the real one, and $c$ is the Fermi
velocity. These fermions obey anticommutation relations: 
$\{\xi_\nu^q(x),\xi_{\nu'}^{q'}(y)\}=\delta_{\nu,\nu'}\delta_{q,q'}\delta(x-y)$, 
$\{\psi_\nu(x),\psi_{\nu'}(y)\}=0$ and 
$\{\psi_\nu(x),\psi_{\nu'}^\dag(y)\}=\delta_{\nu,\nu'}\delta(x-y)$. 
The corresponding
Hamiltonian of the Gaussian theory is 
\begin{eqnarray}
\label{Gau}
\hat{\cal H}[\phi] &=& \int \,dx\,
\frac{c}{2}\left[(\partial_x\theta)^2+(\partial_x\phi)^2\right]
=\int \,dx\,c \left[(\partial_x\phi_L)^2+(\partial_x\phi_R)^2\right],
\nonumber\\ 
\end{eqnarray}
where $\phi=\phi_L+\phi_R$ is the real scalar field, 
$\theta=\phi_L-\phi_R$ is the dual field of $\phi$, and 
$\phi_{L(R)}$ is the left (right) moving part of $\phi$. 
The fields $\phi$ and $\theta$ satisfy the canonical commutation relation
$[\phi(x), \theta(y)]=-\frac{i}{2}{\rm sgn}(x-y)$. Chiral
fields $\phi_{L,R}$ obey $[\phi_L(x),\phi_R(y)]=0$ and 
$[\phi_{L/R}(x),\phi_{L/R}(y)]=\mp\frac{i}{4}{\rm sgn}(x-y)$.
In condensed-matter physics, 
the Hamiltonian~(\ref{Dirac_Majorana}) usually
originates from a microscopic system in solids 
(e.g., a lattice system such as the Hubbard chain and 
the Heisenberg one) after a coarse-graining or
a renormalization procedure.

Among these fermion and boson fields, operator identities hold.
The fermion annihilation (or creation) operators are bosonized as 
\begin{eqnarray}
\label{boso_1}
\psi_L=\frac{\kappa_L}{\sqrt{2\pi\alpha}}\exp(-i\sqrt{4\pi}\phi_L),
\hspace{0.5cm}
\psi_R=\frac{\kappa_R}{\sqrt{2\pi\alpha}}\exp(i\sqrt{4\pi}\phi_R),
\end{eqnarray}
where $\kappa_{L,R}$ are Klein factors which satisfy 
$\{\kappa_\nu,\kappa_{\nu'}\}=2\delta_{\nu,\nu'}$, and are necessary for
the boson vertex (exponential) operators to reproduce 
the correct anticommutation relation between $\psi_L$ and $\psi_R$. 
The parameter $\alpha$ is a short-distance cut off, which depends on
details of the microscopic model considered. (Note that it is possible
to construct another formula without Klein factors, although it requires
a modification of commutation relations among bosons. See
Refs.~\cite{Shan,Delf}.) Following Haldane's
harmonic-fluid approach~\cite{Hal_boson,Gia,H-F}, 
one can obtain an alternative bosonized form of  
$\psi_L$ and $\psi_R$: when a real-space fermion field $\psi(x)$ in the
considering microscopic system is
approximated as $\psi\sim e^{ik_Fx}\psi_R+e^{-ik_Fx}\psi_L$, one may
bosonize $\psi_{L,R}$ as 
\begin{eqnarray}
\label{boso_1_v2}
e^{-ik_Fx}\psi_L=\frac{\kappa_L}{\sqrt{2\pi\alpha}}\sum_{n=0}^{\infty} 
e^{-i(2n+1)(\sqrt{\pi}\phi + k_F x)}e^{-i\sqrt{\pi}\theta},\nonumber\\
e^{ik_Fx}\psi_R=\frac{\kappa_R}{\sqrt{2\pi\alpha}}\sum_{n=0}^{\infty} 
e^{i(2n+1)(\sqrt{\pi}\phi + k_F x)}e^{-i\sqrt{\pi}\theta}.
\end{eqnarray}
The quantity $k_F$ is the Fermi wave number in the microscopic
system. 
%$\rho=\langle \psi_L^\dag\psi_L+\psi_R^\dag\psi_R\rangle$ is the
%total density of the Dirac fermion, which diverges if the band width is
%inifinite. 
The $n=0$ most relevant terms
correspond to Eq.~(\ref{boso_1}).  
The chiral U(1) currents 
${\cal J}_L=:\psi_L^\dag\psi_L:$ and ${\cal J}_R=:\psi_R^\dag\psi_R:$
%where $:\cdots:$ means normal-ordered product,  
are written as 
\begin{eqnarray}
\label{boso_2}
{\cal J}_L=\frac{1}{\sqrt{\pi}}\partial_x\phi_L,
\hspace{1cm}
{\cal J}_R=\frac{1}{\sqrt{\pi}}\partial_x\phi_R.
\end{eqnarray}
In addition to the correspondences between the fermion and the boson, 
it is known~\cite{Gogo,CFT,Shelton,Fu-Zh,Fran,Fab} 
that the Ising order and disorder fields
$\sigma_\nu$ and $\mu_\nu$ can be bosonized as~\cite{note_Klein2} 
\begin{eqnarray}
\label{boso_3}
\sigma_1\sigma_2\sim\sin(\sqrt{\pi}\phi),
\hspace{1cm}
\mu_1\mu_2\sim\cos(\sqrt{\pi}\phi),\nonumber\\
\sigma_1\mu_2\sim\cos(\sqrt{\pi}\theta),
\hspace{1cm}
\mu_1\sigma_2\sim\sin(\sqrt{\pi}\theta).
\end{eqnarray}

Following these Abelian bosonization rules, one can bosonize 1D 
interacting Dirac fermion systems as well as the free massless
fermion~(\ref{Dirac_Majorana}). If the interaction terms are all
irrelevant in the sense of the renormalization group, the effective
Hamiltonian at the low-energy limit 
is still a Gaussian type with the velocity $c$ and 
coefficients of $(\partial_x\phi)^2$ and $(\partial_x\theta)^2$
renormalized. Conventionally~\cite{Gia}, 
the resultant Hamiltonian is written as 
\begin{eqnarray}
\label{renormalized_H}
\hat{\cal H}^{\rm eff}[\phi] &=& \int dx \frac{c'}{2}
\left[{g}(\partial_x{\theta})^2
+\frac{1}{g}(\partial_x{\phi})^2\right],
\end{eqnarray}
where $c'$ is the renormalized velocity, and $g$ is called the TLL
parameter (the Hamiltonian~(\ref{Gau}) corresponds to a $g=1$ theory). 
When a system is reduced to this type at the
low-energy limit, we say that the system belongs to the TLL 
universality~\cite{Hal_boson,Gogo,Tsv,Gia}. The Gaussian
theory~\cite{Gogo,Tsv,Gia,CFT,Mag} yields 
\begin{eqnarray}
\label{corr-func}
\langle\partial_x\phi(x)\partial_x\phi(0)\rangle=-\frac{K}{2\pi}\frac{1}{x^2},
\nonumber\\
\langle e^{iq\sqrt{\pi}\phi(x)}e^{-iq'\sqrt{\pi}\phi(0)}\rangle
\sim\delta_{q,q'}\left(\frac{\alpha}{x}\right)^{q^2g/2},
\nonumber\\
\langle e^{iq\sqrt{\pi}\theta(x)}e^{-iq'\sqrt{\pi}\theta(0)}\rangle
\sim\delta_{q,q'}\left(\frac{\alpha}{x}\right)^{q^2/(2g)},
\end{eqnarray}
where $q$ and $q'$ are an integer (note the
comment~\cite{note_compact}). This result shows that the scaling
dimensions of $\partial_x\phi$, $e^{iq\sqrt{\pi}\phi}$ and 
$e^{iq\sqrt{\pi}\theta}$ are respectively $1$, $q^2g/4$ and
$q^2/(4g)$.

%%%%%%%%%%%%%%%%%%%%%%%%%%%%%%%%%%%%%%%%%%%%%%%%%%%%%%%%
%%%%%%%%%%%%%%%%%%%%%%%%%%%%%%%%%%%%%%%%%%%%%%%%%%%%%%%%
%%%%%%%%%%%%%%%%%%%%%%%%%%%%%%%%%%%%%%%%%%%%%%%%%%%%%%%%
%%%%%%%%%%%%%%%%%%%%%%%%%%%%%%%%%%%%%%%%%%%%%%%%%%%%%%%%

%\setcounter{section}{2}
\section{Correlation functions of massive fermion theories}
\label{FermiCorrlation}
In this Appendix, we evaluate correlation functions of two massive 
fermion systems, 
\begin{eqnarray}
\hat{\cal H}[\xi^3] &=&\int \,dx\,
\frac{i}{2}v_0\left(\xi_L^3\partial_x\xi_L^3-\xi_R^3\partial_x\xi_R^3\right)
+mi\xi_L^3\xi_R^3, \label{fermi1}
\nonumber\\
&=& \sum_k\epsilon_0(k)\tilde\eta_0^\dag(k)\tilde\eta_0(k),\\
\hat{\cal H}[\eta_-] &=&\int\, dx\,\eta_-^\dag
\left(-\frac{v_0^2}{2m}\partial_x^2+m+H+{\cal O}(\partial_x^4)\right)\eta_-
\nonumber\\
&=& \sum_k\epsilon_-(k)\tilde\eta_-^\dag(k)\tilde\eta_-(k),\label{fermi2}
\end{eqnarray}
where $\xi_{L,R}^3$ is the chiral real fermion field, and other fields 
$\eta_-$ and $\tilde\eta_{0,-}$ are defined in Sec.~\ref{inducedTLL}. 

First, we consider the Majorana fermion system~(\ref{fermi1}). 
At $T=0$, $\langle \tilde\eta_0^\dag(k)\tilde\eta_0(k')\rangle=0$ and 
$\langle \tilde\eta_0(k)\tilde\eta_0^\dag(k')\rangle=\delta_{k,k'}$. 
Therefore, the two-point function of $\xi_L^3$ is calculated as 
\begin{eqnarray}
\label{xi_2point}
\langle\xi_L^3(x)\xi_L^3(0)\rangle &=& \frac{1}{L}\sum_{k>0}
\frac{1}{2\epsilon_0(k)}\left[(\epsilon_0(k)-kv_0)e^{ikx}
+(\epsilon_0(k)+kv_0)e^{-ikx}\right]
\nonumber\\
&\approx&\frac{1}{2}\delta(x)+\frac{1}{2}\int_{-\Lambda}^\Lambda 
\frac{dk}{2\pi}\frac{kv_0}{\epsilon_0(k)}e^{-ikx}
\nonumber\\
&\approx& -i\frac{2m}{\pi v_0}K_1(x/\xi_c),
\hspace{0.5cm}({\rm at}\,\,x\gg\xi_c),
\end{eqnarray}
where we used Eqs.~(\ref{Fourier_fermion}), (\ref{Bogoliu}) 
and (\ref{Bogo_matrix}).
Here, $\Lambda$ is the ultraviolet cut off, $\xi_c=v_0/m$, and 
$K_\nu$ is the modified Bessel function 
($K_\nu(z)\sim \sqrt{\frac{2\pi}{z}}e^{-z}$ at $z\gg 1$). 
Similarly, one can obtain
\begin{eqnarray}
\label{xi_2point_part2}
\langle\xi_R^3(x)\xi_R^3(0)\rangle &=&
\langle\xi_L^3(x)\xi_L^3(0)\rangle, 
\nonumber\\
\langle\xi_L^3(x)\xi_R^3(0)\rangle &\approx & -i \frac{m}{2\pi
 v_0}K_0(x/\xi_c), \hspace{1cm}({\rm at} \,\,x\gg \xi_c). 
\end{eqnarray}

In another system~(\ref{fermi2}), the similar relations
$\langle \tilde\eta_-^\dag(k)\tilde\eta_-(k')\rangle=0$ and 
$\langle \tilde\eta_-(k)\tilde\eta_-^\dag(k')\rangle=\delta_{k,k'}$ hold at $T=0$.
One hence easily finds 
\begin{eqnarray}
\label{eta_2point}
\langle\eta_-^\dag(x)\eta_-(0)\rangle &=& 0,
\nonumber\\
\langle\eta_-(x)\eta_-^\dag(0)\rangle &=& \delta(x)\sim\frac{\delta_{x,0}}{\alpha},
\end{eqnarray}
where $\alpha$ is the short-distance cut off.

In addition to these results, one can of course compute 
any correlation functions of the systems~(\ref{fermi1}) and
(\ref{fermi2}), using Wick's theorem, etc.

%%%%%%%%%%%%%%%%%%%%%%%%%%%%%%%%%%%%%%%%%%%%%%%%%%%%%%%%%%%%
%%%%%%%%%%%%%%%%%%%%%%%%%%%%%%%%%%%%%%%%%%%%%%%%%%%%%%%%%%%%
%%%%%%%%%%%%%%%%%%%%%%%%%%%%%%%%%%%%%%%%%%%%%%%%%%%%%%%%%%%%
%%%%%%%%%%%%%%%%%%%%%%%%%%%%%%%%%%%%%%%%%%%%%%%%%%%%%%%%%%%%
%%%%%%%%%%%%%%%%%%%%%%%%%%%%%%%%%%%%%%%%%%%%%%%%%%%%%%%%%%%%
%%%%%%%%%%%%%%%%%%%%%%%%%%%%%%%%%%%%%%%%%%%%%%%%%%%%%%%%%%%%

\section{Symmetries of the spin-1/2 XXZ chain}
\label{XXZ}
The spin-1/2 XXZ chain~(\ref{xxz}) has the same global symmetries 
as those of the spin-1 AF chain~(\ref{spin1AF}): 
the U(1) rotation around the spin z axis, the
one-site translation, and the site-parity transformation.  
In the Abelian bosonization framework (see Eqs.~(\ref{Gaussian}) and
(\ref{spin_xxz})), 
these three symmetries could be realized by the following
transformations of boson fields $\bar\phi$ and $\bar\theta$~\cite{Aff-Lec,Egg}. 

(\rnum{1}) The U(1) rotation $S_j^+\to e^{i\gamma}S_j^+$ corresponds to 
\begin{eqnarray}
\label{U(1)_xxz}
\bar\theta \to \bar\theta +\frac{\gamma}{\sqrt{\pi}}.
\end{eqnarray}

(\rnum{2}) The one-site translation $\vec S_j\to \vec S_{j+1}$ corresponds to 
\begin{eqnarray}
\label{onesite_xxz}
\bar\phi(x) &\to& \bar\phi(x+a_0)+\sqrt{\pi}\left(\bar M+\frac{1}{2}\right),
\nonumber\\
\bar\theta(x) &\to& \bar\theta(x+a_0)+\sqrt{\pi}.
\end{eqnarray}

(\rnum{3}) The site-parity transformation $\vec S_j\to \vec S_{-j}$ corresponds to 
\begin{eqnarray}
\label{parity_xxz}
\bar\phi(x) &\to& -\bar\phi(-x)+\frac{\sqrt{\pi}}{2},
\nonumber\\
\bar\theta(x) &\to& \bar\theta(-x).
\end{eqnarray}

%%%%%%%%%%%%%%%%%%%%%%%%%%%%%%%%%%%%%%%%%%%%%%%%%%%%%%%%%%%%
%%%%%%%%%%%%%%%%%%%%%%%%%%%%%%%%%%%%%%%%%%%%%%%%%%%%%%%%%%%%
%%%%%%%%%%%%%%%%%%%%%%%%%%%%%%%%%%%%%%%%%%%%%%%%%%%%%%%%%%%%
%%%%%%%%%%%%%%%%%%%%%%%%%%%%%%%%%%%%%%%%%%%%%%%%%%%%%%%%%%%%
%%%%%%%%%%%%%%%%%%%%%%%%%%%%%%%%%%%%%%%%%%%%%%%%%%%%%%%%%%%%
%%%%%%%%%%%%%%%%%%%%%%%%%%%%%%%%%%%%%%%%%%%%%%%%%%%%%%%%%%%%


\begin{thebibliography}{99}
\bibitem{radius}The relationship between our notation using $\bar K$ and
	another using a compactification radius $R$ for the field
	$\bar\phi$ is given by $\bar K=1/(4\pi R^2)$. 
	In addition to these two, there are other normalization forms 
	in the Abelian bosonization. See for instance Ref.~\cite{Gia}.
\bibitem{note1}A part of the bosonized-spin expression has been already
	completed. It however contains only a few relevant terms. 
	See Refs.~\cite{Aff-bose,Aff-magBEC,Kon,Fath,Ess-Aff}.
\bibitem{note2}We used Eqs. (2.31a) and (2.31b) in Ref.~\cite{Wu}, and
	the asymptotic expansion formula of modified Bessel function, 
	$K_\nu(z)\sim\sqrt{\frac{\pi}{2z}}e^{-z}[1+(4\nu^2-1)/(8z)
	+(4\nu^2-1)(4\nu^2-3)/(2!(8z)^2)+\cdots]$.
\bibitem{note_compact}In the standard compactified boson theory (see
	Refs.~\cite{Gogo,Tsv,CFT,Aff-Lec,Mag}) as an
	effective theory for a lattice system, the coexistence
	of $e^{\pm i\sqrt{\pi}\Phi}$ and $e^{\pm i\sqrt{\pi}\Theta}$,
	which are present in Eqs.~(\ref{spin-uniform}) and (\ref{spin-stagg}), is
	prohibited, because the former vertex operator requires 
	the compactification radius $R_\Phi=1/\sqrt{\pi}$ for the scalar field
	$\Phi$, but the latter does $R_\Phi=1/\sqrt{4\pi}$. However,
	since the present theory consists of the boson part and the
	remaining Ising theory $(\xi^3,\sigma^3,\mu^3)$, we think that
	the ``entanglement'' between these two parts (i.e., three copies
	of Ising systems) allows such a coexistence. 
	In other words, the field $\Phi$ is expected not to be
	simply compactified, as long as we concurrently consider both the
	boson and the Ising parts.
\bibitem{note_Klein}Imposing a nontrivial transformation on Klein
	factors might be strange. 
	Therefore, there is the possibility that we can find 
	an alternating, more suitable transformation.
\bibitem{note_MS}There are a few mistakes in the calculation of $\vec
	J_L\vec J_R$ in Section IV C of Ref.~\cite{MS05_1}, although
	they do not change the main results in Ref.~\cite{MS05_1}.
\bibitem{note_CumExp}A systematic method of the integration over
	the massive fields is the cumulant expansion for the 
	partition function. For example, refer to Wang Y -J, Essler F H
	L, Fabrizio M and Nersesyan A A, 2002 {\it Phys. Rev. B} {\bf
	66}, 024412. For the present effective theory, the cumulant
	expansion would be considered as a
	$\lambda_{\parallel,\perp}/v$ expansion. Thus, it would be valid as
	far as the Fermi velocity $v(\sim v_F)$ is sufficiently large, i.e., $H$ is
	sufficiently larger than the Haldane gap $m$.
\bibitem{note_theta}The emergence of $(\partial_x\theta)^2$ might be a
	failure in the field theory strategy. See Sections 6.1.2 and
	7.2.1 in Ref.~\cite{Gogo}.
\bibitem{note_product}We should carefully treat a product between two
	continuous fields that coordinates are very close each other
	[e.g., $e^{i\sqrt{\pi}\Theta(x)}\times \psi_L(x)$ 
	in Eq.~(\ref{vertex_fermion_1})], 
	because such fields possess no information on the scale of the
	lattice constant $a_0$. Therefore, prefactors in all terms 
	of Eqs.~(62) and (63) %~(\ref{vertex_fermion_1})
	would not be reliable at least quantitatively.
\bibitem{note_Commute}Note that $\zeta_{L/R}$ and $\psi_{L/R}$ commute 
	with each other in Eq.~(62). Thus, we should interpret that 
	$\zeta_{L/R}$ and $\kappa_{L/R}$ (or $\eta_-$) also commute with
	each other.
\bibitem{note_Klein2}In order to more accurately treat the commutation
	relations among vertex operators, fermions $\xi_\nu^q$ 
	and Ising fields ($\sigma^q$,$\mu^q$), one had better insert
	a certain quantity such as Klein factors in
	Eq.~(\ref{boso_3}). For example see Ref.~\cite{Fab}.
\end{thebibliography}
\end{document}